\documentclass{article}

\usepackage{arxiv}

\usepackage[utf8]{inputenc} 
\usepackage[T1]{fontenc}    
\usepackage{hyperref}       
\usepackage{url}            
\usepackage{booktabs}       
\usepackage{nicefrac}       
\usepackage{microtype}      
\usepackage{lipsum}

\usepackage{pgfplots}
\usepackage{amsmath,amsfonts,amssymb,amsthm}
\usepackage{subcaption}
\usepackage{mathtools}
\usepackage{commath}
\usepackage{nicefrac}
\usepackage{booktabs}
\usepackage{textcomp}
\usepackage{cite}
\usepackage{adjustbox}

\title{NVIDIA SimNet\texttrademark: an AI-accelerated multi-physics simulation~framework}

\author{
  \textcolor{black}{Oliver Hennigh} \\
  \texttt{ohennigh@nvidia.com} \\
  \And
  Susheela Narasimhan \\
  \texttt{susheelan@nvidia.com}
  \And
  \textcolor{black}{Mohammad Amin Nabian} \\
  \texttt{mnabian@nvidia.com} \\
  \And
  Akshay Subramaniam \\
  \texttt{asubramaniam@nvidia.com} \\
  \And
  \textcolor{black}{Kaustubh Tangsali} \\
  \texttt{ktangsali@nvidia.com} \\
  \And
  \textcolor{black}{Max Rietmann} \\
  \texttt{mrietmann@nvidia.com}
  \And
  Jose del Aguila Ferrandis \\
  \texttt{jaguila@mit.edu} \\
  \And
  Wonmin Byeon \\
  \texttt{wbyeon@nvidia.com} \\
  \And
  Zhiwei Fang \\
  \texttt{zhiweif@nvidia.com} \\
  \And
  Sanjay Choudhry\thanks{Corresponding author} \\
  \texttt{schoudhry@nvidia.com}
}

\begin{document}
\maketitle

\begin{abstract}
We present SimNet, an AI-driven multi-physics simulation framework, to accelerate simulations across a wide range of disciplines in science and engineering. Compared to traditional numerical solvers, SimNet addresses a wide range of use cases - coupled forward simulations without any training data, inverse and data assimilation problems. SimNet offers fast turnaround time by enabling parameterized system representation that solves for multiple configurations simultaneously, as opposed to the traditional solvers that solve for one configuration at a time. SimNet is integrated with parameterized constructive solid geometry as well as STL modules to generate point clouds. Furthermore, it is customizable with APIs that enable user extensions to geometry, physics and network architecture. It has advanced network architectures that are optimized for high-performance GPU computing, and offers scalable performance for multi-GPU and multi-Node implementation with accelerated linear algebra as well as FP32, FP64 and TF32 computations. In this paper we review the neural network solver methodology, the SimNet architecture, and the various features that are needed for effective solution of the PDEs. We present real-world use cases that range from challenging forward multi-physics simulations with turbulence and complex 3D geometries, to  industrial design optimization and inverse problems that are not addressed efficiently by the traditional solvers. Extensive comparisons of SimNet results with open source and commercial solvers show good correlation.
\end{abstract}


\section{Introduction}

Simulations are pervasive in every domain of science and engineering. However, they become computationally expensive as more geometry details are included and as model size, the complexity of physics or the number of design evaluations increases. Since large simulations can sometimes take hours to days to complete a single run, the speed of iteration in the workflow is constrained by the speed of simulations. Although deep learning \cite{lecun2015deep} offers a path to overcome this constraint, supervised learning techniques are used most often in the form of traditional data driven neural networks (e.g., \cite{guo2016convolutional,kim2020prediction,hennigh2017lat}). However, generating data can be an expensive and time consuming process. Furthermore, these models may not obey the governing physics of the problem, involve extrapolation and generalization errors, and provide unreliable results.

In comparison with the traditional solvers, neural network solvers \cite{raissi2019physics,lagaris1998artificial,sirignano2018dgm} can not only do parametrized simulations in a single run, but also address problems not solvable using traditional solvers, such as inverse or data assimilation problems and real time simulation. They can also be embedded in the traditional solvers to improve the predictive capability of the solvers. In its most basic form, a neural network solver consists of a standard fully-connected neural network or multi-layer perceptron. The loss functions used to train these networks are augmented by Partial Differential Equations (PDEs) describing a physical process, and require computing the derivatives of the outputs with respect to the inputs. Initial and boundary conditions can be imposed as hard constraints by changing the network architecture or more generally, used in the loss function to fully specify the physical constraints. Training of neural network forward solvers can be supervised based on the governing laws of a physics only, and thus, unlike the data-driven deep learning models, neural network solvers do not require any training data. However, for data assimilation or inverse problems, data constraints are introduced in the loss function. Automatic differentiation is used to compute the derivatives required for the residuals of the PDEs. Gradients of the loss function are then back-propagated through the entire network for training using a stochastic gradient descent optimizer. 

\begin{figure}[htp]
    \centering
    \includegraphics[width=1.0\textwidth]{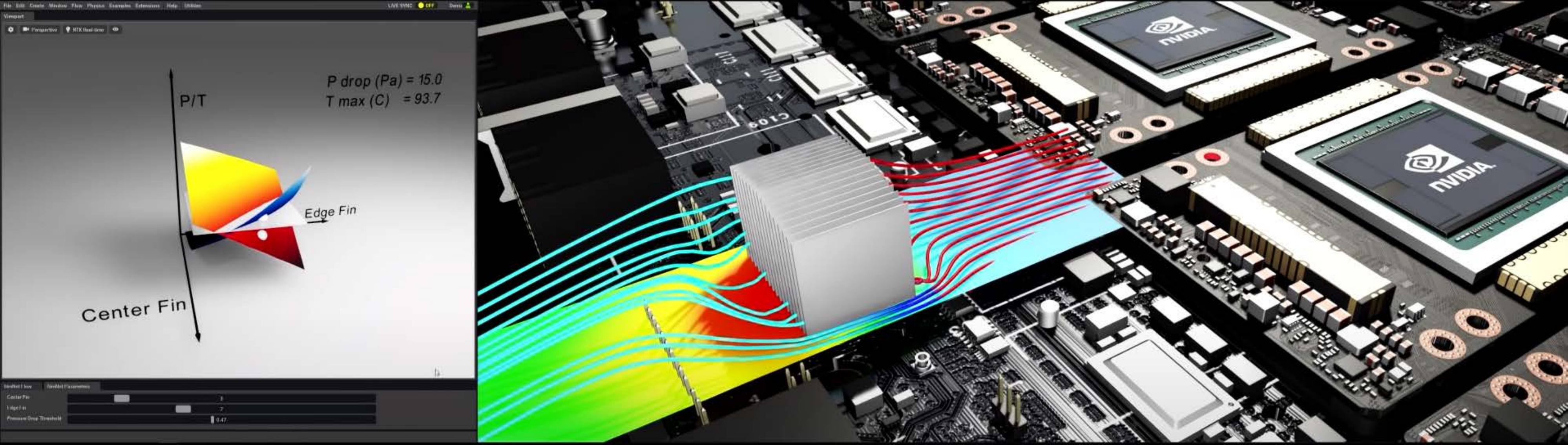}
    \caption{Design optimization of an FPGA heat sink using SimNet. The center and side fin heights are the two design variables, with a linear transition for the height of the intermediate fins.}
    \label{fig:fpga_design_optimization}
\end{figure}

Rapid evolution of GPU architecture suited for AI and HPC, as well as introduction of open source frameworks like Tensorflow have motivated researchers to develop novel algorithms for solving PDEs (e.g., \cite{wang2020and,wang2020understanding,berg2018unified,jagtap2020adaptive,kharazmi2020hp,peng2020accelerating,zang2020weak,raissi2020hidden,michoski2020solving,haghighat2020deep,zhu2019physics,jin2020nsfnets,shin2020convergence}). Recently, a number of neural network solver libraries are being developed aiming at making these solvers more accessible to the academia and researchers. Among those are Tensorflow-based DeepXDE \cite{lu2019deepxde}, Keras-based SciANN \cite{haghighat2020sciann}, and Julia-based NeuralPDE.jl \cite{DifferentialEquations.jl-2017}. Although the existing research studies and libraries played a crucial role in advancing the neural network solvers, the attempted examples are mostly limited to simple 1D or 2D domains with straightforward governing physics, and the neural network solvers in their current form still struggle to forward solve real-world applications that involve complex 3D geometries and multi-physics systems. In this paper we present SimNet, a new framework for academia as well as industry, that aims to address the current computational challenges with neural network solvers. As an example, SimNet enables design optimization of a FPGA heat sink (see Figure \ref{fig:fpga_design_optimization}) through a single network training without any training data. In contrast, the traditional solvers are not capable of simulating geometries with several design parameters in a single run. 

\vspace{-5mm}
\textcolor {black} {\paragraph{Our Contributions:} Several research studies have recently been published demonstrating solution of PDEs using neural networks. However, our experience has shown that they do not converge well when used as forward solvers for industrial problems due to the gradients, singularities and discontinuities introduced by complex geometries or physics. Our main contributions in this paper are to offer several novel features to address these challenges - Signed Distance Functions (SDFs) for loss weighting, integral continuity planes for flow simulation, advanced neural network architectures, point cloud generation for real world geometries using constructive geometry module as well as STL module and finally parameterization of both geometry and physics. Additionally, for the first time to our knowledge, we solve high Reynolds number flows using zero-equation turbulence model in industrial applications without using any data.}

\section{AI driven Simulations}

\begin{figure}[htp]
    \centering
    \includegraphics[width=0.8\textwidth]{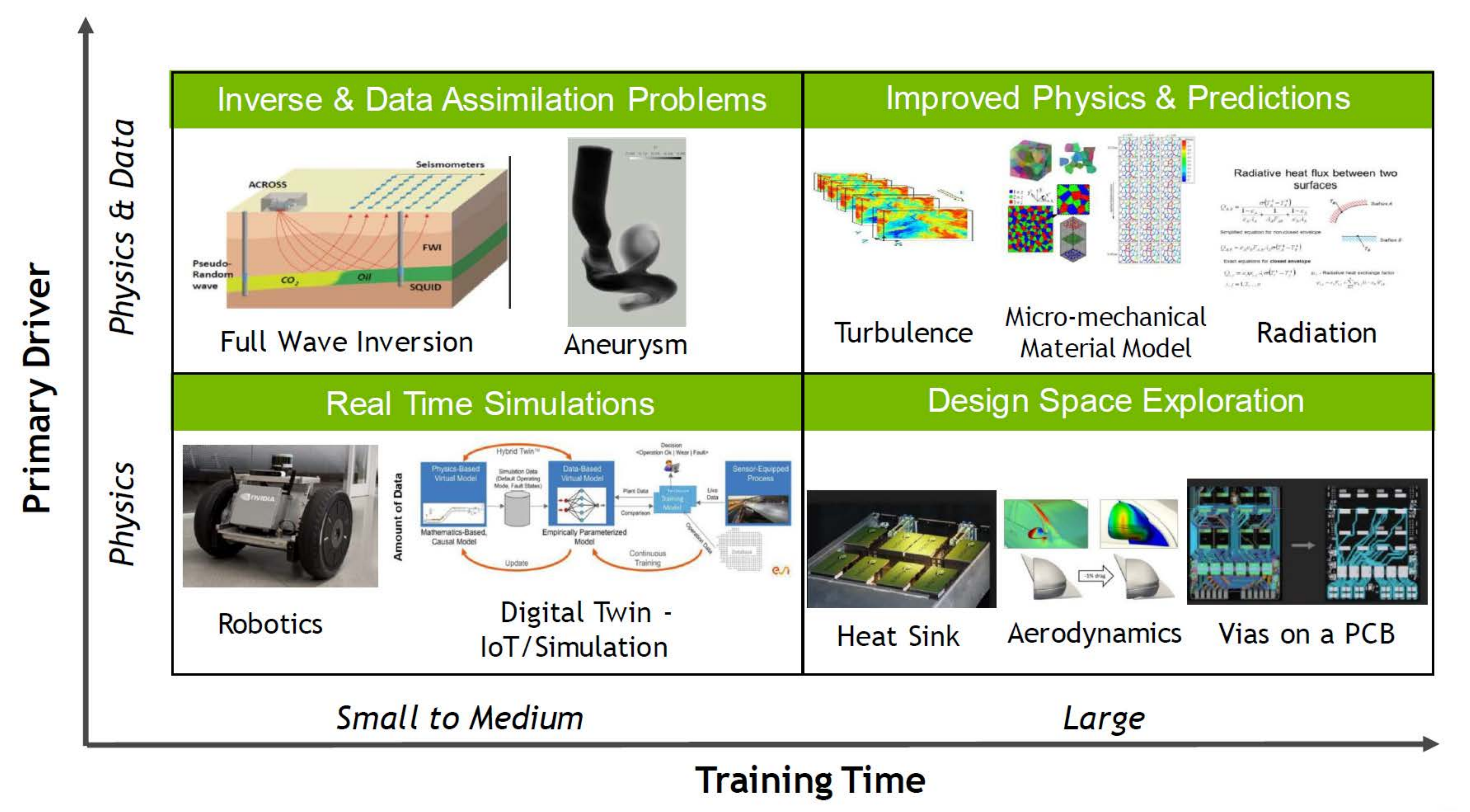}
    \caption{Four major areas in computational science and engineering addressed by SimNet.}
    \label{fig:applications}
\end{figure}

Neural network solvers are capable of addressing the various areas in the sciences and engineering as shown in Figure \ref{fig:applications}. Other areas such as uncertainty quantification and sensitivity analysis could also benefit from the neural network solvers.

\vspace{-2mm}
\paragraph{Design Space Exploration and Optimization:} Accurate physical simulation of products, processes and systems saves time and cost in product development by providing engineers with feedback on their designs without building and testing physical prototypes. In a fast-paced competitive world where manufacturers are constantly trying to reduce the time-to-market and cost while optimizing the product behavior and manufacturing processes, companies are turning increasingly towards simulations. However, due to time and computational resource constraints, only a limited number of simulations can be performed during a design cycle. Neural network solvers provide a fast simulation surrogate to traditional simulations that allows fast design space exploration due to ability to work with several different geometries simultaneously as well as parameterized design for a single geometry.

\vspace{-2mm}
\paragraph{Improved physics and predictions:} Traditional solvers use various models to represent physical phenomenon. These models can be replaced by neural network solvers trained on highly accurate physics results or experiments. For example, viscosity and turbulence models can be learned using the neural networks on high fidelity results (e.g. DNS) and experiments. These learnt models can then be called from a computationally inexpensive framework (e.g. RANS) from within a solver to get a higher fidelity at a lower computational expense. Besides fluid mechanics, examples of other physics areas that can benefit from such modeling are radiation in heat transfer and micromechanics of materials in solids. 

\vspace{-2mm}
\paragraph{Inverse Modeling:} Neural network solvers are a natural choice to be used for inverse problems where the underlying physical behavior is discovered by assimilating data from observations. The physics can range from acoustics with applications in oil and gas exploration (e.g., using recorded sound data to find the sub-surface velocity model) to fluid mechanics with applications in medical imaging (e.g. given a 4D CT or MR images of intracranial aneurysm or stenosis in the artery, one can back calculate various physics quantities such as velocities and pressure using a passive scalar). Neural network solvers also present an attractive alternative to traditional simulations where the right boundary conditions are not easily available.

\vspace{-2mm}
\paragraph{Real-Time Simulations:} Several situations require nearly instantaneous feedback where simulations would provide valuable information but are not possible due to the real-time nature of the situation. Examples include robots, digital twins, autonomous driving and real time evaluation of aerodynamics to influence driving decisions on race-tracks. Neural network solvers provide the capability for real-time simulations in these systems. A pre-trained neural network solver can provide real-time feedback in a matter of seconds, and thus, can make a huge impact in the areas that real-time simulations are crucial.

The rest of the paper is structured as follows. In Section \ref{sec:method}, we provide a brief overview on the theoretical aspects of the neural network solvers. The architecture of the SimNet is then introduced next in Section \ref{sec:simnet}. The capability of SimNet in solving complex real-world problems involving turbulent multi-physics simulations, complex geometries, design optimization, and inverse simulations is illustrated in Section \ref{sec:use_cases}. We discuss performance improvements in Section \ref{HPC}, and conclude with a summary of the main contributions of SimNet.

\section{Neural Network Solvers} \label{sec:method}
\begin{figure}[htp]
    \centering
    \includegraphics[width=0.85\textwidth]{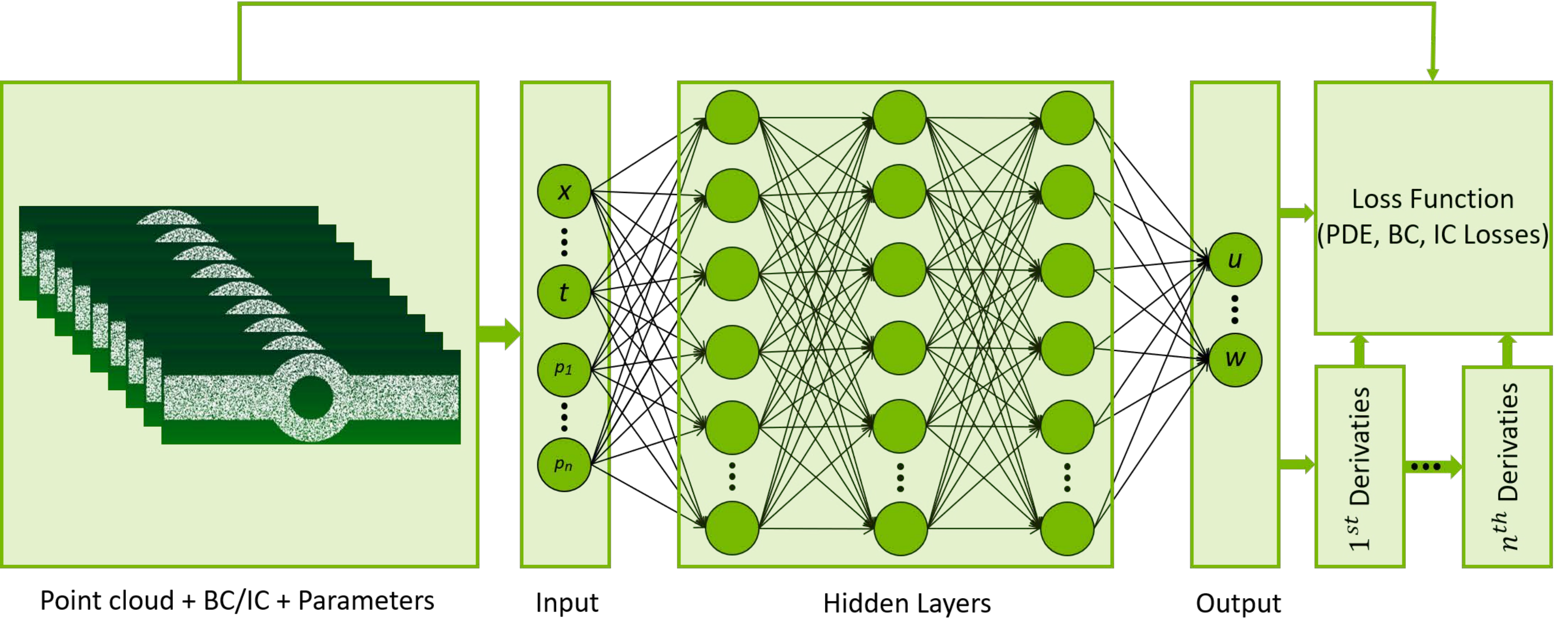}
    \caption{\textcolor{black}{A schematic of the structure of a neural network solver. The inputs to the network are the spatial coordinates of a point cloud, realizations of time (if applicable), and realizations from the parametric space (if applicable). The inputs are mapped to the quantities of interest via a fully-connected network with nonlinear activation functions. To train this network, a loss function is considered which consists of the derivatives of the output w.r.t inputs (computed using automatic differentiation), and initial and boundary condition information.}}
    \label{fig:neural_network_solver}
\end{figure}

In this section we provide an introduction to neural network solvers. Briefly, a neural network solver approximates the solution to a given PDE and a set of boundary and initial constraints using a feed-forward fully-connected neural network. The model is trained by constructing a loss function for how well the neural network is satisfying the PDE and constraints. If the network is able to minimize this loss function then it will in effect, solve the given PDE. A schematic of the structure of a neural network solver is shown in Figure \ref{fig:neural_network_solver}.

Let us consider the following general form of a PDE:

\begin{equation}\label{eqn:pde}
\begin{aligned}
\mathcal{N}_i[u]\left(\mathbf{x}\right) = f_i\left(\mathbf{x}\right), \; \; \; \;  & \forall i \in \{1, \cdots, N_\mathcal{N}\}, \mathbf{x}\in \mathcal{D},
\\
\mathcal{C}_j[u]\left(\mathbf{x}\right) = g_j\left(\mathbf{x}\right), \; \; \; \;  & \forall j \in \{1, \cdots, N_\mathcal{C}\}, \mathbf{x}\in \mathcal{\partial D}, \\
\end{aligned}
\end{equation}

where $\mathcal{N}_i$'s are general differential operators, $\mathbf{x}$ is the set of independent variables defined over a bounded continuous domain $\mathcal{D} \subseteq \mathbb{R}^D , D \in \left \{ 1,2,3,\cdots \right \}$, and $u (\mathbf{x})$ is the solution to the PDE. $\mathcal{C}_j$'s denote the constraint operators that may consist of differential, linear, and nonlinear terms and usually cover the boundary and initial conditions. $\mathcal{\partial {D}}$ also denotes a subset of the domain boundary that is required for defining the constraints. We seek to approximate the solution $u (\mathbf{x})$ by a neural network $u_{net}(\mathbf{x})$ that, in it's most simple form, takes the following form:

\begin{equation}\label{eqn:u_net}
    u_{net}(\mathbf{x};\mathbf{\theta}) = \mathbf{W}_n \big \{\phi_{n-1} \circ \phi_{n-2} \circ \cdots \circ \phi_1 \circ \phi_E \big \} (\mathbf{x}) + \mathbf{b}_n,  \; \; \; \; \phi_{i}(\mathbf{x}_i) = \sigma \left( \mathbf{W}_i \mathbf{x}_i + \mathbf{b}_i \right),
\end{equation}

where $\mathbf{x} \in \mathbb{R}^{d_0}$ is the input to network, $\phi_{i} \in \mathbb{R}^{d_i}$ is the $i^{th}$ layer of the network, $\mathbf{W}_i \in \mathbb{R}^{d_i \times d_{i-1}}, \mathbf{b}_i \in \mathbb{R}^{d_i}$ are the weight and bias of the $i^{th}$ layer, $\mathbf{\theta}$ denotes the set of network's trainable parameters, i.e., $\mathbf{\theta} = \{\mathbf{W}_1, \mathbf{b}_1, \cdots, \mathbf{b}_n, \mathbf{W}_n\}$, $n$ is the number of layers, and $\sigma$ is the activation function. We suppose that this neural network is infinitely differentiable, i.e. $u_{net} \in C^{\infty}$. $\phi_E$ is an input encoding layer, and by setting that to identity function, we arrive at the standard feed-forward fully-connected architecture, which is the most widely used architecture in neural network solvers. More advanced architectures will be introduced in Section \ref{architectures}. 

In order to train this neural network, we construct a loss function that penalizes over the divergence of the approximate solution $u_{net} (\mathbf{\theta})$ from the PDE in equation \ref{eqn:pde}, and such that the constraints are encoded as penalty terms. To this end, we define the following residuals by using $u_{net}$ as the approximate solution to the PDE:

\begin{equation}\label{eqn:res}
\begin{aligned}
r_\mathcal{N}^{(i)} \left(\mathbf{x}; u_{net}(\mathbf{\theta})\right) = \mathcal{N}_i[u_{net}(\mathbf{\theta})]\left(\mathbf{x}\right) - f_i\left(\mathbf{x}\right) ,
\\
r_\mathcal{C}^{(j)} \left(\mathbf{x}; u_{net}(\mathbf{\theta})\right) = \mathcal{C}_j[u_{net}(\mathbf{\theta})]\left(\mathbf{x}\right) - g_j\left(\mathbf{x}\right),
\end{aligned}
\end{equation}

where $r_\mathcal{N}^{(i)}$ and $r_\mathcal{C}^{(j)}$ are the PDE and constraint residuals, respectively. The loss function then takes the following form:


\begin{equation} \label{sumation_loss}
  \mathcal{L}_{res} (\mathbf{\theta}) = \sum_{i=1}^{N_\mathcal{N}} \int_\mathcal{D} \lambda_{\mathcal{N}}^{(i)}(\mathbf{x}) \norm{r_\mathcal{N}^{(i)} \big( \mathbf{x}; u_{net}(\mathbf{\theta}) \big)}_p d \mathbf{x} + \sum_{j=1}^{N_\mathcal{C}} \int_{\partial\mathcal{D}} \lambda_{\mathcal{C}}^{(j)}(\mathbf{x}) \norm{r_\mathcal{C}^{(j)} \big( \mathbf{x}; u_{net}(\mathbf{\theta}) \big)}_p d \mathbf{x},
\end{equation}
where $\norm{\cdot}_p$ denotes the p-norm, and  $\lambda_{\mathcal{N}}^{(i)}, \lambda_{\mathcal{C}}^{(j)}$ are weight functions that control the loss interplay between within and across different terms. The majority of the research studies focus on finding dynamic weight to scale the different loss terms (e.g., \cite{wang2020understanding,wang2020and,jin2020nsfnets,chen2018gradnorm}). Currently, SimNet supports three loss weighting algorithms, that are global learning rate annealing as proposed in \cite{wang2020understanding}, a new local variant of this algorithm, and also signed distance weighting that will be explained in Section \ref{sec:geometry}.

Finally, to train the approximate solution $u_{net} (\mathbf{\theta})$, the network parameters $\theta$ are optimized iteratively using variants of the stochastic gradient descent method, such as the Adam optimizer \cite{kingma2014adam}. At each iteration, the integral terms in the loss function are approximated using a regular or Quasi-Monte Carlo method, and using a batch of samples from the independent variables $\mathbf{x}$. Automatic differentiation is commonly used to compute the required gradients in $\nabla{\mathcal{L}_{res} (\mathbf{\theta})}$.

\section{SimNet Overview} \label{sec:simnet}

\begin{figure}[htp]
    \centering
    \includegraphics[width=0.65\textwidth]{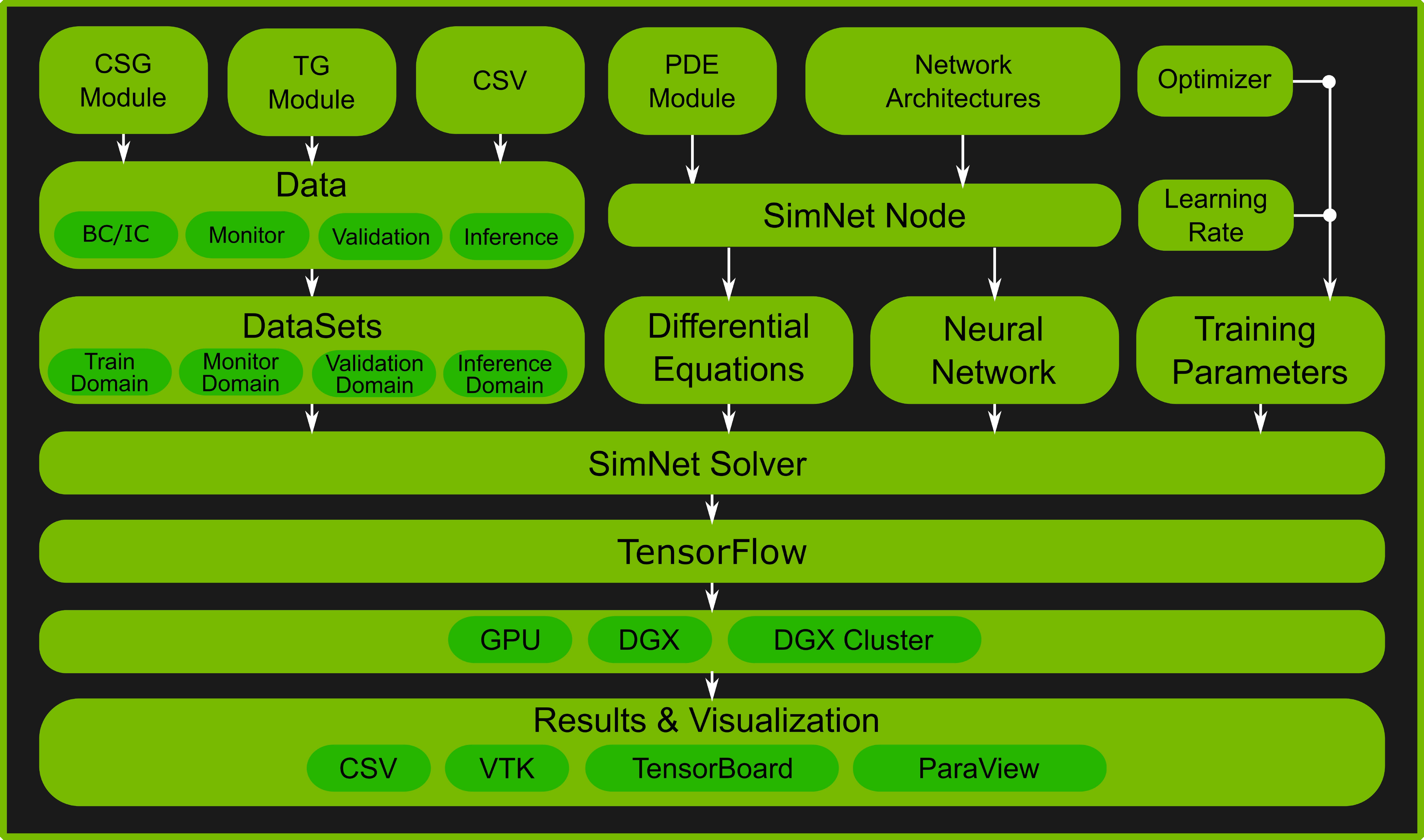}
    \caption{SimNet structure.}
    \label{fig:simnet_structure}
\end{figure}

SimNet is a Tensorflow based neural network solver \cite{abadi2016tensorflow} with source code  available at:  \href{http://developer.nvidia.com/simnet}{developer.nvidia.com/simnet}. It offers various APIs that enable the user to leverage the existing functionality to build their own applications on the existing modules. An overview of SimNet architecture is presented in Figure \ref{fig:simnet_structure}. The geometry modules, PDE module, and data are used to fully specify the physical system. The user also specifies the network architecture, optimizer and learning rate schedule. SimNet then constructs the neural network solver, forms the loss function, and unrolls the graph efficiently to compute the gradients. The SimNet solver then starts the training or inference procedure using TensorFlow's built-in functions on a single or cluster of GPUs. The outputs are saved in form of CSV or VTK files and can be visualized using TensorBoard and ParaView. In the remainder of this section, we present more detail on the SimNet's geometry and PDE modules as well as the available network architectures. 

\subsection{Geometry modules} \label{sec:geometry}
\paragraph{CSG and TG modules:} SimNet currently carries two geometry modules, namely the Constructive Solid Geometry (CSG) and Tessellated Geometry (TG) modules. With SimNet's CSG module, constructive geometry primitives can be easily defined and Boolean operations can be performed. This will allow creation and parameterization of a wide range of geometries and enables definition of various shapes that can be parameterized and used for design optimization. The TG module in SimNet imports STL, OBJ, and other tessellated geometries to work with complex geometries. 

\paragraph{Signed Distance Functions (SDFs):} One area of considerable interest is how to weight the loss terms in the overall loss function, as defined in equation \ref{sumation_loss}. \textcolor{black}{SimNet offers spatial loss weighting, where each weight parameter can be a function of the spatial inputs. In many cases we use the Signed Distance Function (SDF) for this weighting.} Assuming $\mathcal{D}_x$ represents the spatial subset of the input domain $\mathcal{D}$ with boundaries $\partial \mathcal{D}_x \subseteq \partial \mathcal{D}$, the SDF-based weight function is defined as

\begin{equation}\label{eqn:sdf}
    \lambda(\mathbf{x_s}) = \begin{cases}
\,\,\,\, \,d(\mathbf{x_s}, \partial \mathcal{D}_x) & \mathbf{x_s} \in \mathcal{D}_x, \\ 
-d(\mathbf{x_s}, \partial \mathcal{D}_x) & \mathbf{x_s} \in \mathcal{D}_x^c.
\end{cases}
\end{equation}
Here, $\mathbf{x}_s$ is the spatial inputs, and $d(\mathbf{x_s}, \partial \mathcal{D}_x)$ represents the Euclidean distance between $\mathbf{x}_s$ and it's nearest neighbor on $\mathcal{D}_x$. In general, we have found it beneficial to weight losses lower on sharp gradients or discontinuous areas of the domain. For example, if there are discontinuities in the boundary conditions we may have the loss decay set to zero on these discontinuities. Equation residuals can also be weighted by the SDF of the geometries. If the geometry has sharp corners this often results in sharp gradients in the solution of the PDE. Weighting by the SDF tends to mitigate the deleterious effects of sharp local gradients, and often results in a convergence speed increase as well as improved accuracy in some cases. \textcolor{black}{To accelerate the computation of the SDF on tessellated meshes of complex geometries, we developed a custom library that leverages NVIDIA's OptiX for both inside/outside (sign) testing and distance computation. The sign test uses ray intersection and triangle normal alignment (via dot product). The distance testing is done by using the bounded volume hierarchy (BVH) interface provided by OptiX, which yields excellent performance and accuracy for distance computations.}

\paragraph{Point cloud generation:} Training points in SimNet are generated according to a uniform distribution by default, which then enable regular Monte Carlo approximation of the loss function in equation \ref{sumation_loss}. Alternatively, SimNet's CSG module also enables quasi-random point cloud generation using the generalized Halton sequences, which provides the means to generate training points with a low level of discrepancy across the domain to perform quasi-Monte Carlo approximation of the loss function. SimNet also has a continuous point sampling algorithm which achieves similar accuracy as the fixed point cloud without the cost of generation.

\subsection{PDE module} \label{PDE}

The PDE module in SimNet contains of a variety of common differential equations including the Navier-Stokes, diffusion, advection-diffusion, wave equations, \textcolor{black}{and linear elasticity equations}. In order to make the PDE module extensible for the user to easily define their own differential equations, SimNet uses symbolic mathematics enabled by SymPy \cite{meurer2017sympy}. The PDE module in SimNet also provides implementations of turbulence and exact continuity in the Navier-Stokes equations:

\paragraph{Zero-equation turbulence model:} Currently, SimNet adopts the zero equation turbulence model for turbulent physics simulation. The zero-equation turbulence model is defined as:
\begin{equation}\label{zeroeq1}
\mu_t (\mathbf{x})=\rho l_m^2 \left(G(\mathbf{x})\right)^\frac{1}{2},
\end{equation}
\begin{equation}\label{zeroeq2}
l_m (\mathbf{x}) =\min \big( 0.419 d \left(\mathbf{x},\partial \mathcal{D}_x\right), 0.09d_{max} \big),
\end{equation}
\begin{equation}\label{zeroeq3}
\resizebox{1.0\hsize}{!}{$
G(\mathbf{x})=2 \bigg[ \left(\frac{\partial u}{\partial x} (\mathbf{x})\right) ^2 + \left(\frac{\partial v}{\partial y} (\mathbf{x})\right) ^2 + \left(\frac{\partial w}{\partial z} (\mathbf{x})\right) ^2 \bigg] + \left(\frac{\partial u}{\partial y} (\mathbf{x}) + \frac{\partial v}{\partial x} (\mathbf{x})\right)^2
+ \left(\frac{\partial u}{\partial z} (\mathbf{x}) + \frac{\partial w}{\partial x} (\mathbf{x})\right)^2 +
\left(\frac{\partial v}{\partial z} (\mathbf{x_s}) + \frac{\partial w}{\partial y} (\mathbf{x})\right)^2.
$}
\end{equation}
Here, $\mu_t \left(\mathbf{x}\right) = \rho \nu_t \left(\mathbf{x}\right)$, $l_m$ is the mixing length, $G$ is the modulus of mean squared rate of strain tensor, $d$ represents the normal distance from wall, and $d_{max}$ is maximum normal distance. The zero-equation turbulence model requires normal distance and its spatial derivatives from no slip walls to compute the turbulent viscosity, and is computed using the SDF from the CSG or TG modules. More advanced turbulence models can be implemented in the current framework.

\paragraph{Exact continuity and integral continuity for incompressible Navier-Stokes equations:} 
Velocity-pressure formulations are the most widely used formulations of the Navier-Stokes equations. Alternatively, one can ensure exact mass conservation using the velocity field from a vector potential \cite{young2015novel}, defined as:
\begin{equation}\vec{V}=\nabla \times \vec{\psi}=\left(\frac{\partial \psi_{z}}{\partial y}-\frac{\partial \psi_{y}}{\partial z}, \frac{\partial \psi_{x}}{\partial z}-\frac{\partial \psi_{z}}{\partial x}, \frac{\partial \psi_{y}}{\partial x}-\frac{\partial \psi_{x}}{\partial y}\right)^{T},\end{equation}
where $\vec{\psi}=\left(\psi_{x}, \psi_{y}, \psi_{z}\right)$. This definition of the velocity field ensures that it is divergence-free and that it satisfies continuity, that is:
\begin{equation}\nabla \cdot \vec{V}=\nabla \cdot(\nabla \times \vec{\psi})=0.\end{equation} 
Although exact continuity is effective in improving the convergence and accuracy, it increases the model training time as it will introduce third-order differential terms. Alternatively, in addition to solving the Navier-Stokes equations in differential form, specifying the volumetric flow rate through some integral continuity planes that are located on the outlet and across the channel significantly speeds up the convergence rate and improves accuracy.

\subsection{Network architectures} \label{architectures}
In addition to the feed-forward, fully connected networks, SimNet offers a number of more advanced architectures, out of which three of the most effective ones are introduced here. Moreover, SimNet offers a large array of activation functions, including the adaptive activation functions proposed in \cite{jagtap2020adaptive}.

\textcolor{black}{\paragraph{Fourier feature networks:} Neural networks are generally biased toward low-frequency solutions, a phenomenon that is known as "spectral bias" \cite{rahaman2019spectral}. This can adversely affect the training convergence and accuracy of the model. One approach to alleviate this issue is to perform input encoding, that is, to transform the inputs to a higher-dimensional feature space via high-frequency functions \cite{mildenhall2020nerf,rahaman2019spectral,tancik2020fourier}. The Fourier feature network in SimNet is a variation of the one proposed in \cite{tancik2020fourier} with trainable encoding, and takes the form in equation \ref{eqn:u_net} with the following encoding
\begin{equation}\label{eqn:fourier_feature}
\phi_E = \big[ \sin \left( 2\pi \mathbf{f} \times \mathbf{x} \right); \cos \left( 2\pi \mathbf{f} \times 
\mathbf{x} \right) \big]^T,
\end{equation}
where $\mathbf{f} \in \mathbb{R}^{n_f \times d_0}$ is the trainable frequency matrix and $n_f$ is the number of frequency sets.}

\textcolor{black}{\paragraph{Modified Fourier feature networks:} The modified Fourier feature network is SimNet's novel architecture, where two transformation layers are introduced to project the Fourier features to another learned feature space, and are then used to update the hidden layers through element-wise multiplications, similar to its standard fully connected counterpart in \cite{wang2020understanding}. It is shown in the next section that this multiplicative interaction can improve the training convergence and accuracy. The hidden layers in this architecture take the following form
\begin{equation}\label{eqn:u_net}
\phi_{i}(\mathbf{x}_i) = \left(1 - \sigma \left( \mathbf{W}_i \mathbf{x}_i + \mathbf{b}_i \right) \right) \odot \sigma \left( \mathbf{W}_{T_1} \phi_E + \mathbf{b}_{T_1} \right) + \sigma \left( \mathbf{W}_i \mathbf{x}_i + \mathbf{b}_i \right) \odot \sigma \left( \mathbf{W}_{T_2} \phi_E + \mathbf{b}_{T_2} \right),
\end{equation}
where $i>1$ and $\{ \mathbf{W}_{T_1}, \mathbf{b}_{T_1}\}, \{ \mathbf{W}_{T_2}, \mathbf{b}_{T_2}\}$ are the parameters for the two transformation layers. The multiplicative interactions between the Fourier features and hidden layers can potentially improve the model's convergence and accuracy, although at the cost of slightly increasing the training time per iteration.}

\paragraph{SiReNs:} The authors in \cite{sitzmann2020implicit} proposed a neural network using Sin activation functions dubbed Sinusoidal Representation Networks or SiReNs. This network has similarities to the Fourier feature networks above because using a Sin activation function has the same effect as the input encoding for the first layer of the network. A key component of this network architecture is the initialization scheme. The weight matrices of the network are drawn from a uniform distribution $\mathbf{W}_i \sim U(-\sqrt{{6/d_{i-1}}},\sqrt{{6/d_{i-1}}})$. The input of each Sin activation has a Gauss-Normal distribution and the output of each Sin activation, an arcSin distribution. This preserves the distribution of activations allowing deep architectures to be constructed and trained effectively \cite{sitzmann2020implicit}. The first layer of the network is scaled by a factor $\omega$ (with a default value of 30) to span multiple periods of the Sin function and empirically shown to give good performance.

\section{Use Cases} \label{sec:use_cases}
In this section, we present four use cases for SimNet to illustrate its capabilities. These use cases are turbulent and multi-physics simulation, simulation with complex geometries, design optimization for a multi-physics system, and an inverse problem. \textcolor{black}{Although, SimNet is capable of simulating transient flows using the continuous-time sampling approach \cite{raissi2019physics}, the use cases presented here are time-independent. A more efficient and accurate approach based on the convolutional LSTMs for transient simulations is under development}.

For the entire neural network solvers in this section, the architectures consist of 6 layers, each with 512 units, and Swish \cite{ramachandran2017searching} nonlinearities. For the simulations presented in Sections \ref{example2} to \ref{example4}, the standard fully connected architecture is used. We use the Adam optimizer with an initial learning rate of $10^{-4}$, an exponential decay, and the TensorFlow's default values for the other optimizer parameters. In training our neural network solvers, we use the regular Monte Carlo integration scheme for the loss function in equation \ref{sumation_loss}. Moreover, for the 3D channel flow problems, we use integral continuity planes. For the simulations in use cases \ref{example1} to \ref{example3}, we use the SDF for weighting the PDE residuals.
It must be noted that use cases \ref{example1} to \ref{example3}, are solved in the forward manner without using any training data. 

\subsection{Turbulent and multi-physics simulations} \label{example1}
In this part, using an FPGA heat sink example, we demonstrate the SimNet's capability in accurately solving multi-physics problems involving high Reynolds number flows. The geometry of the FPGA heat sink placed inside a channel is depicted in Figure \ref{fig:fpga}, and the details of the problem setup are reported in Appendix \ref{appendix:fpga}. This particular geometry is challenging to simulate due to thin fin spacing that causes sharp gradients which are particularly difficult to learn for a neural network solver. Using the zero-equation turbulence model, we will solve a conjugate heat transfer problem to simulate flow over the heat sink placed inside a channel at $Re=13,239.6$. Since we are dealing with an incompressible flow, there is a one-way coupling between the heat and flow equations, and it is possible to train the temperature field after the flow field is trained. To this end, we train two separate neural network solvers, one for the flow field and the other one for the temperature field. This approach is useful for one-way coupled multi-physics problems as it is possible to achieve significant speed-up, and also simulate cases with same flow boundary conditions but different thermal boundary conditions. For instance, one can easily use the same flow field as in input to train for different thermal boundary conditions. 

\begin{figure}[htp]
\centering
\begin{subfigure}{0.25\textwidth}
\includegraphics[width=1\textwidth]{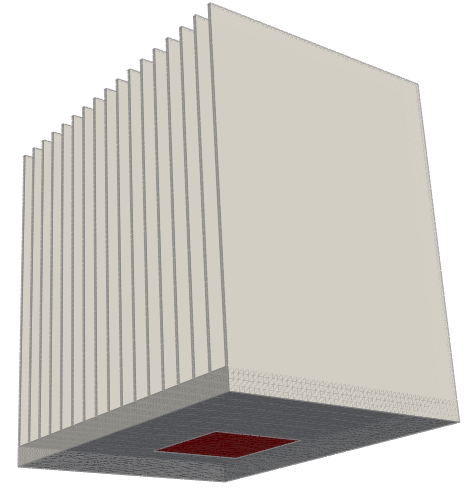}
\caption{}
\label{fig:fpga}
\end{subfigure} 
\hspace{0.02\textwidth}
\begin{subfigure}{0.32\textwidth}
\includegraphics[width=1\textwidth]{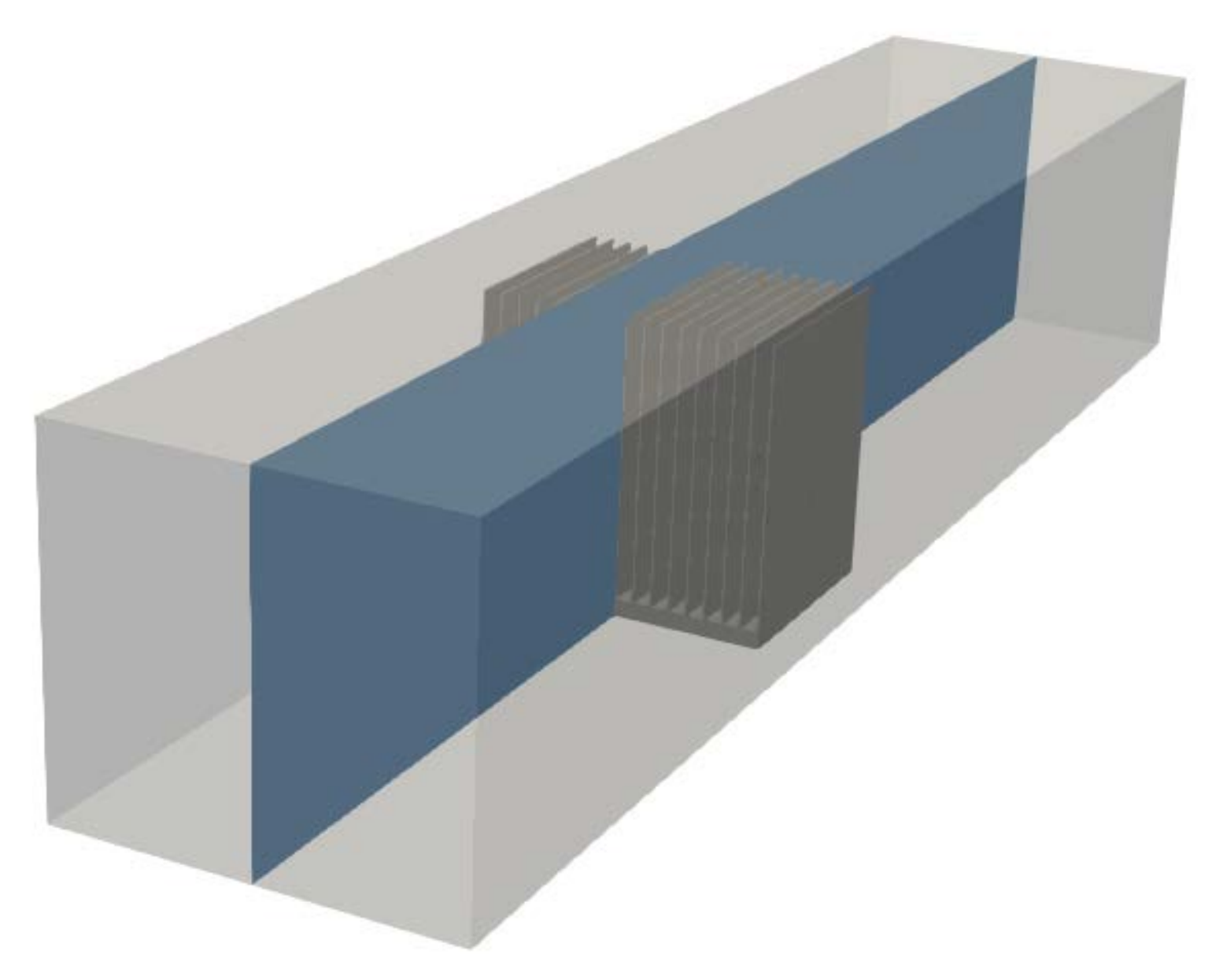}
\caption{}
\label{fig:fpga_channel}
\end{subfigure}
\caption{FPGA heat sink example. (a) The FPGA heat sink geometry; (b) the simulation domain. The blue plane represents the plane of symmetry.}
\label{fig:fpga}
\end{figure}

We leverage symmetry of the problem to reduce the computational domain, accelerate the training, reduce the memory usage, and potentially, improve the accuracy. To this end, we utilize the following symmetry boundary conditions at the plane of symmetry, as shown in Figure \ref{fig:fpga_channel}: (1) zero value for the physical variables that are odd functions w.r.t. the plane of symmetry, and (2) zero normal gradient for physical variables that are even functions w.r.t. the plane of symmetry.

\begin{figure}[htp]
\centering
\begin{subfigure}{0.49\textwidth}
\includegraphics[width=1\textwidth]{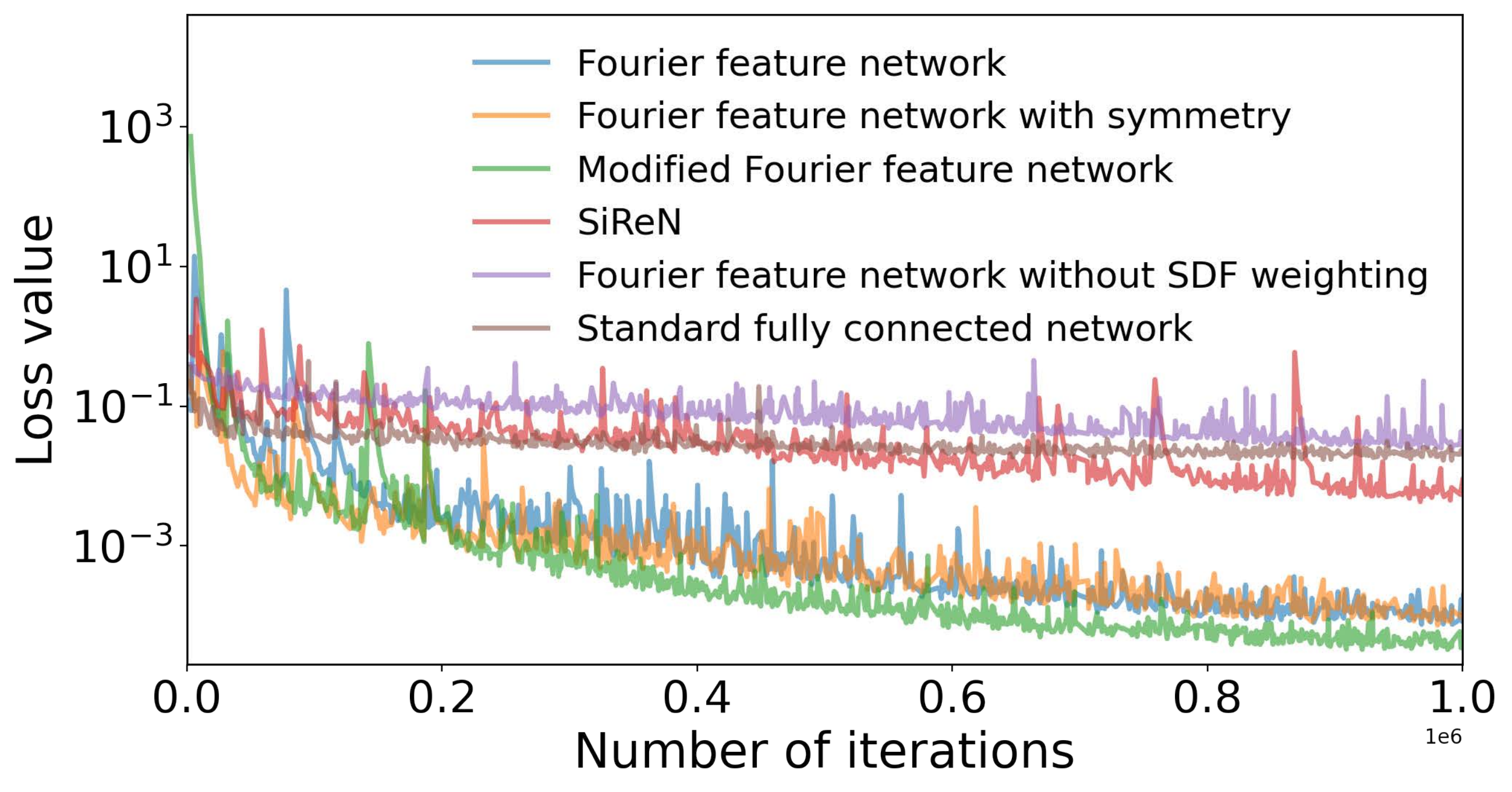}
\caption{FPGA flow training}
\end{subfigure} 
\begin{subfigure}{0.49\textwidth}
\includegraphics[width=1\textwidth]{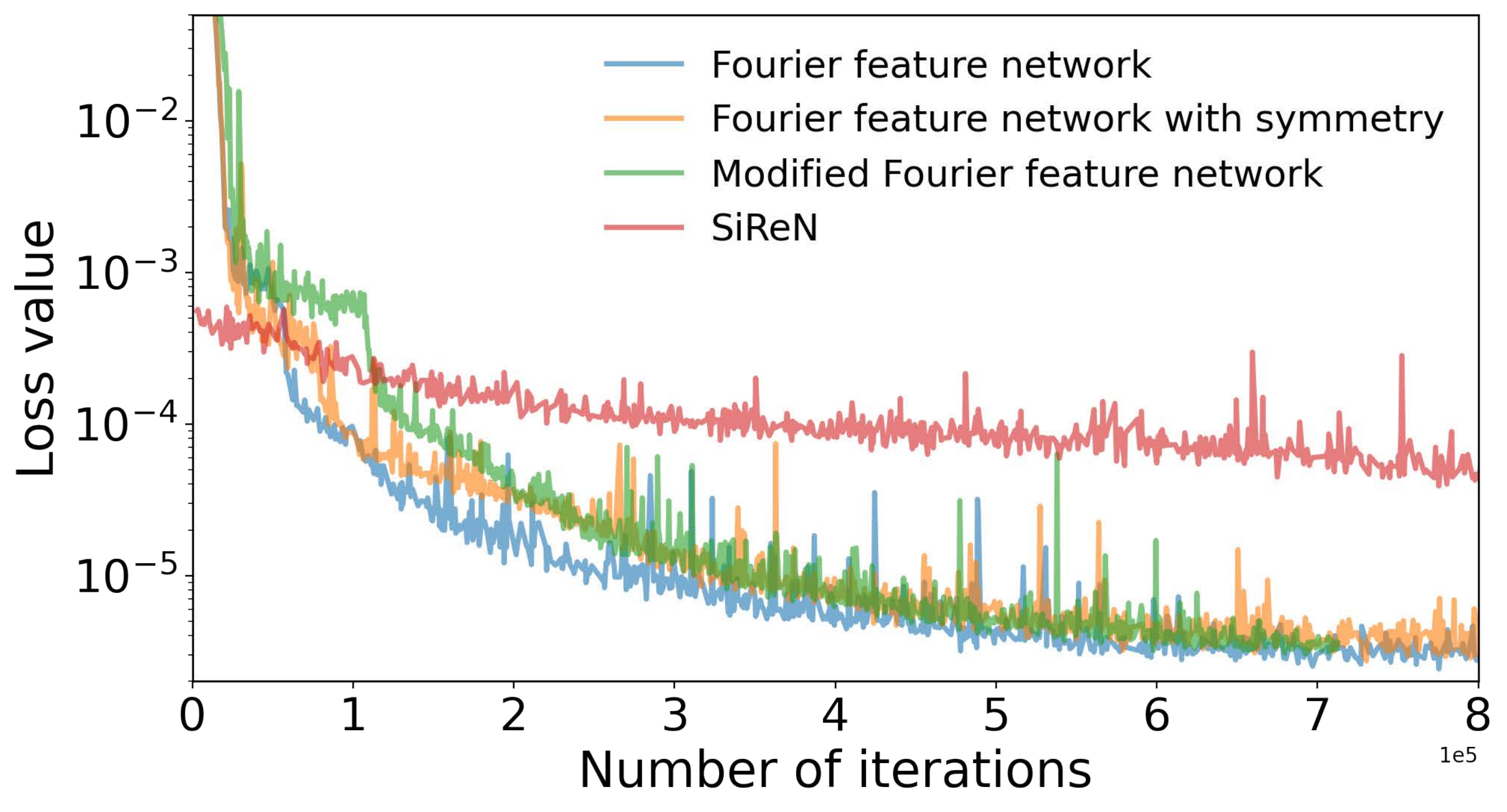}
\caption{FPGA heat training}
\end{subfigure}
\caption{Loss curves for the training of the FPGA conjugate heat transfer problem using different network architectures.}
\label{fig:fpga_loss}
\end{figure}

\begin{figure}[htp]
    \centering
    \includegraphics[width=0.55\textwidth]{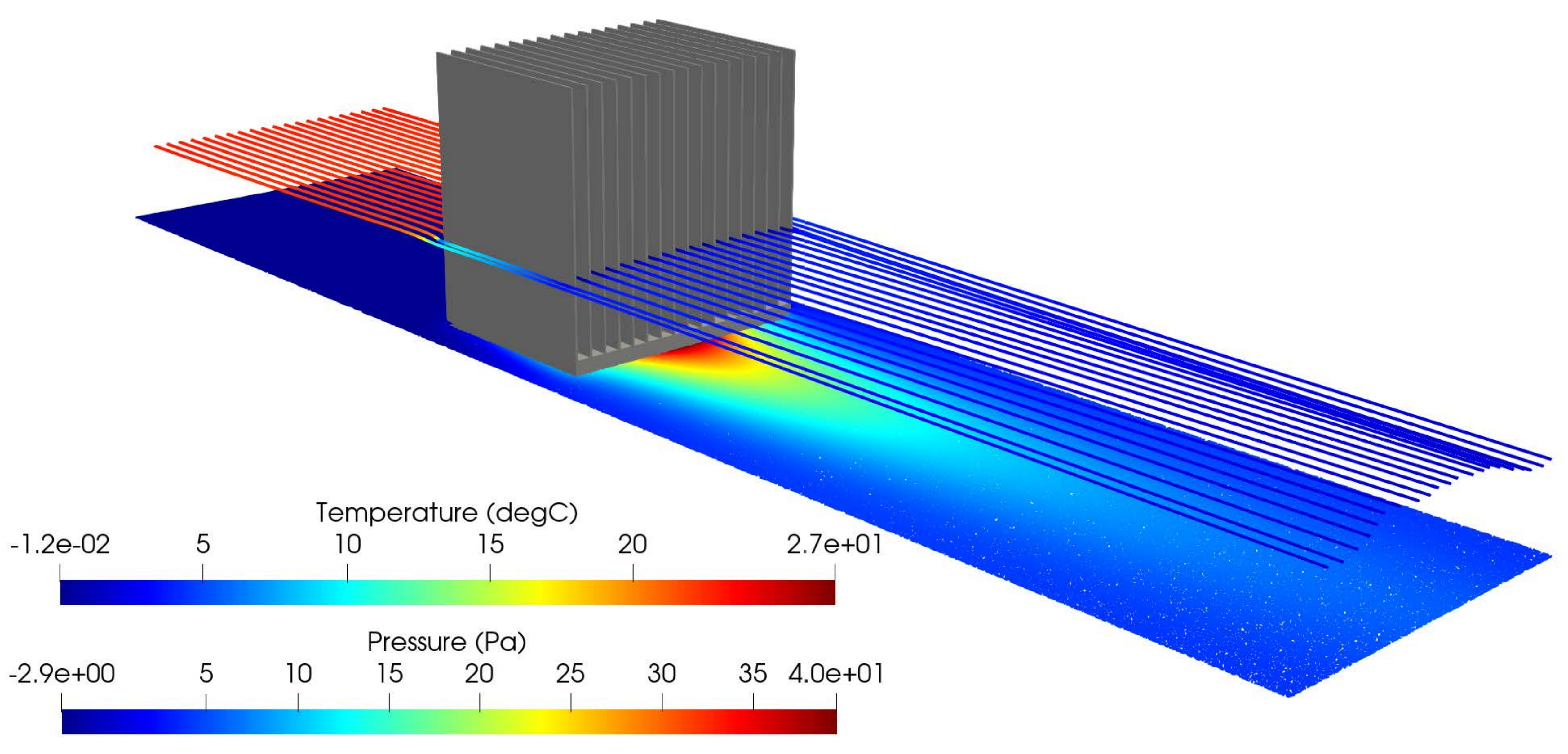}
    \caption{Streamlines and temperature profile obtained from the SimNet model for FPGA with modified Fourier feature network.}
    \label{fig:fpga_streamlines}
\end{figure}

\begin{figure}[!h]
\centering
\begin{subfigure}{0.32\textwidth}
\includegraphics[width=1\textwidth]{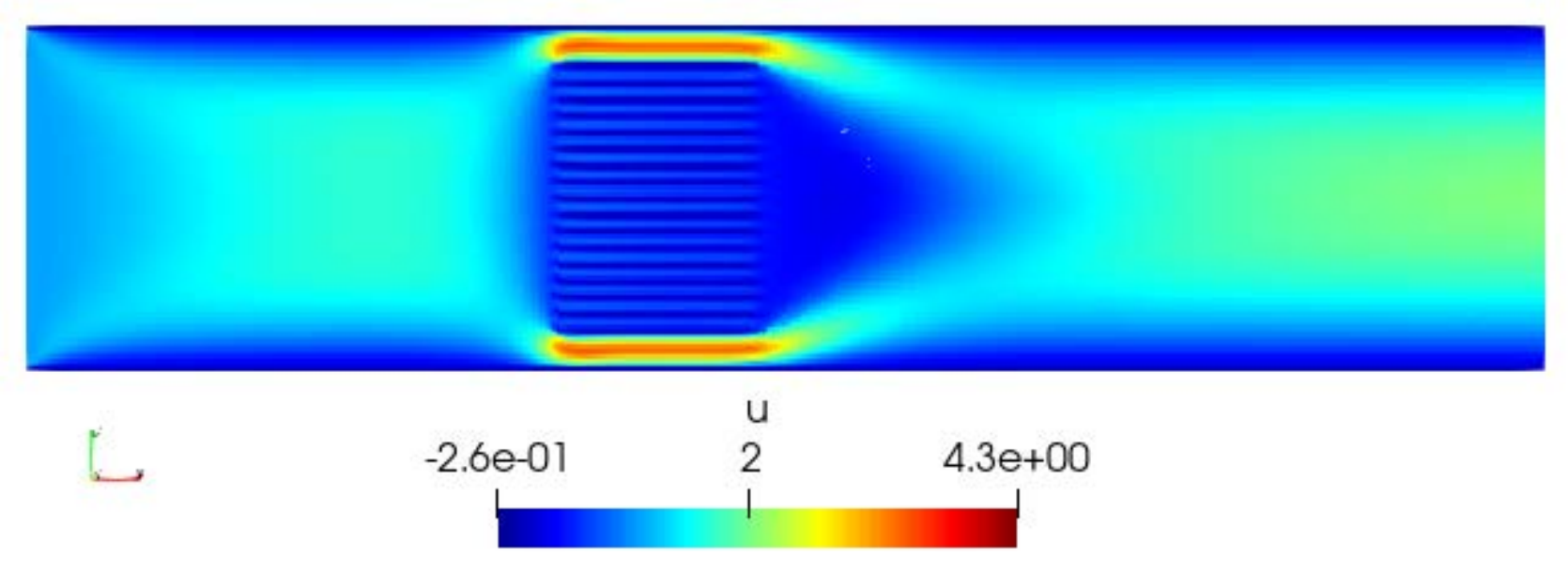}
\caption{$u$-velocity (SimNet)}
\end{subfigure} 
\begin{subfigure}{0.32\textwidth}
\includegraphics[width=1\textwidth]{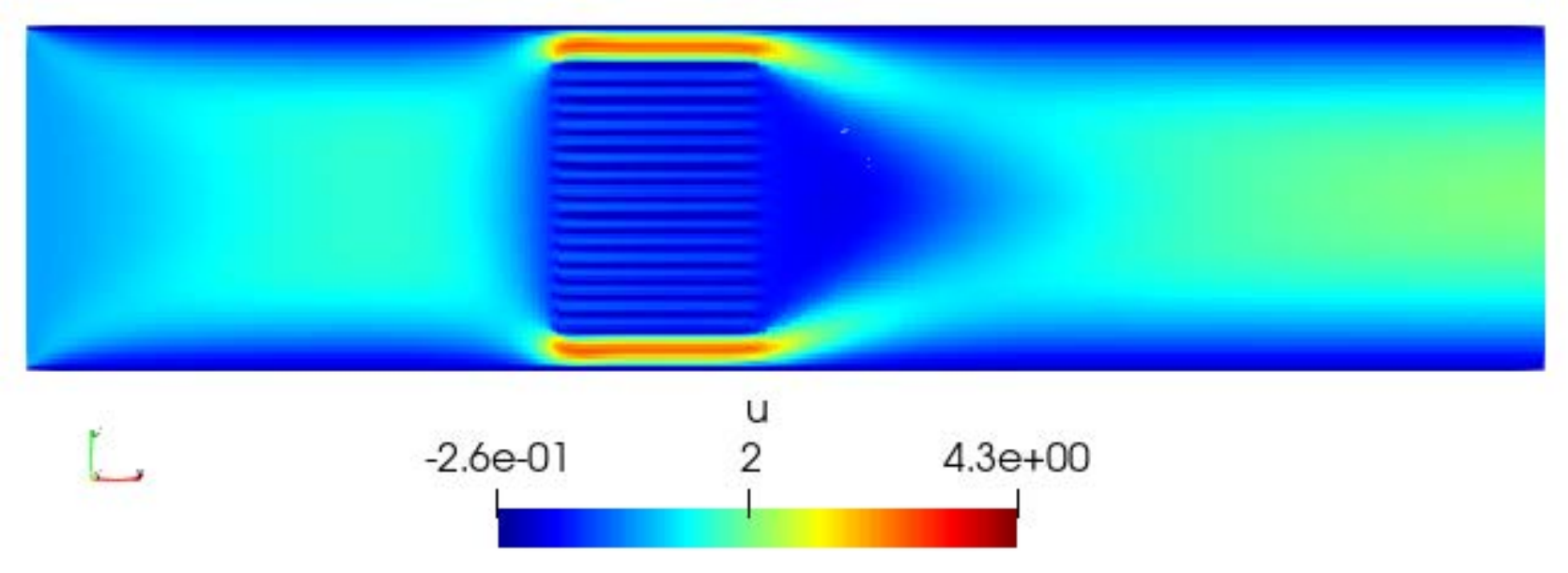}
\caption{$u$-velocity (OpenFOAM)}
\end{subfigure}
\begin{subfigure}{0.32\textwidth}
\includegraphics[width=1\textwidth]{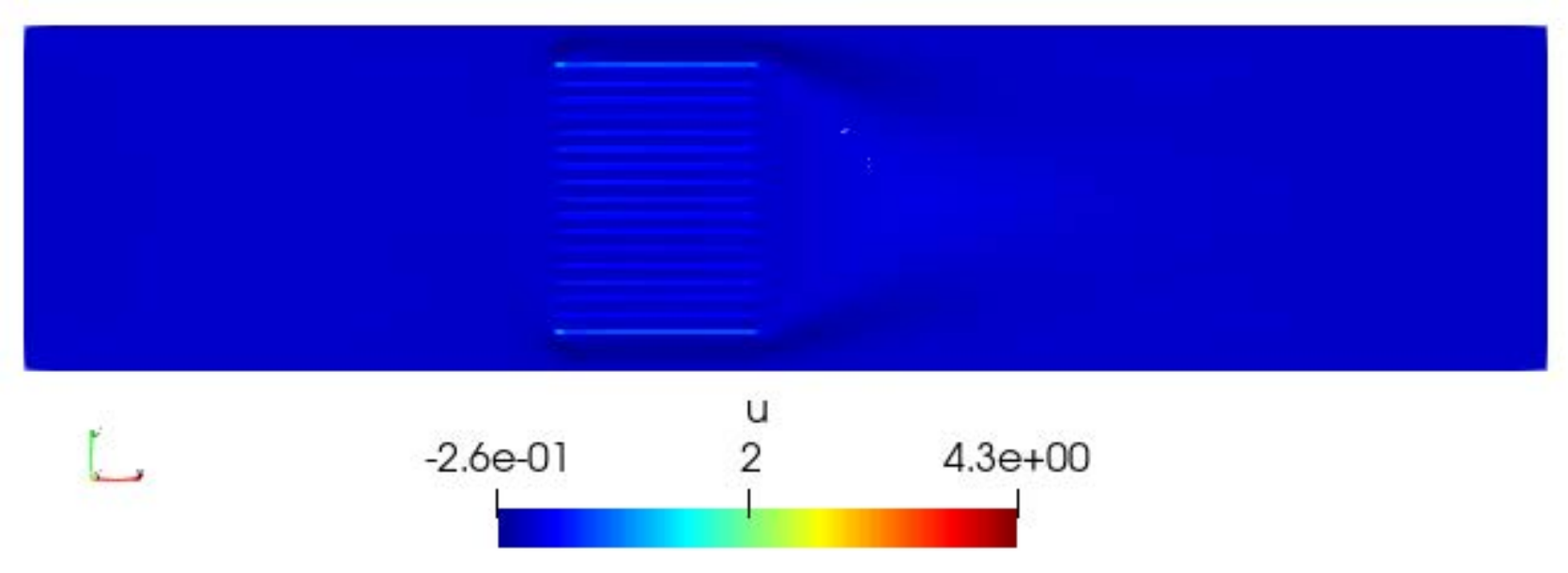}
\caption{$u$-velocity (Difference)}
\end{subfigure} 
\\
\begin{subfigure}{0.32\textwidth}
\includegraphics[width=1\textwidth]{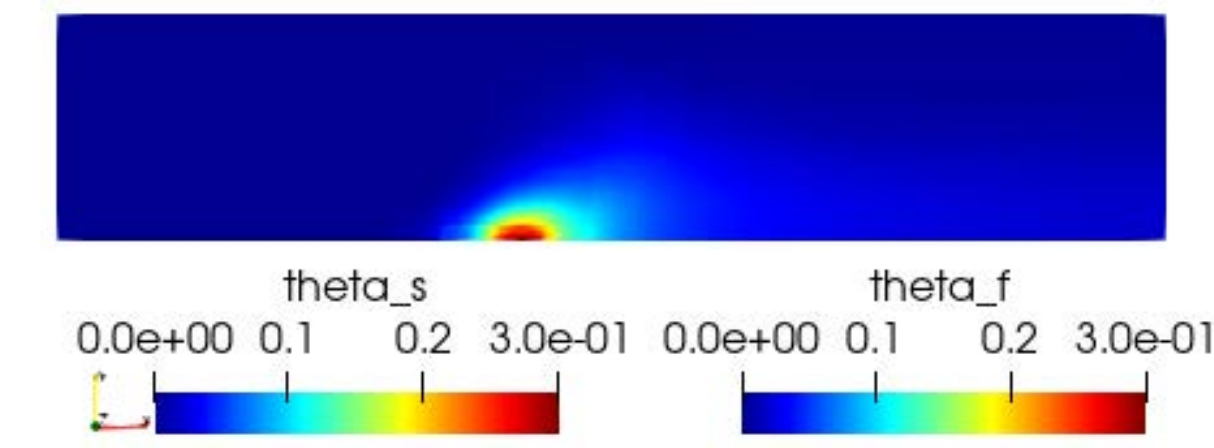}
\caption{Temperature (SimNet)}
\end{subfigure} 
\begin{subfigure}{0.32\textwidth}
\includegraphics[width=1\textwidth]{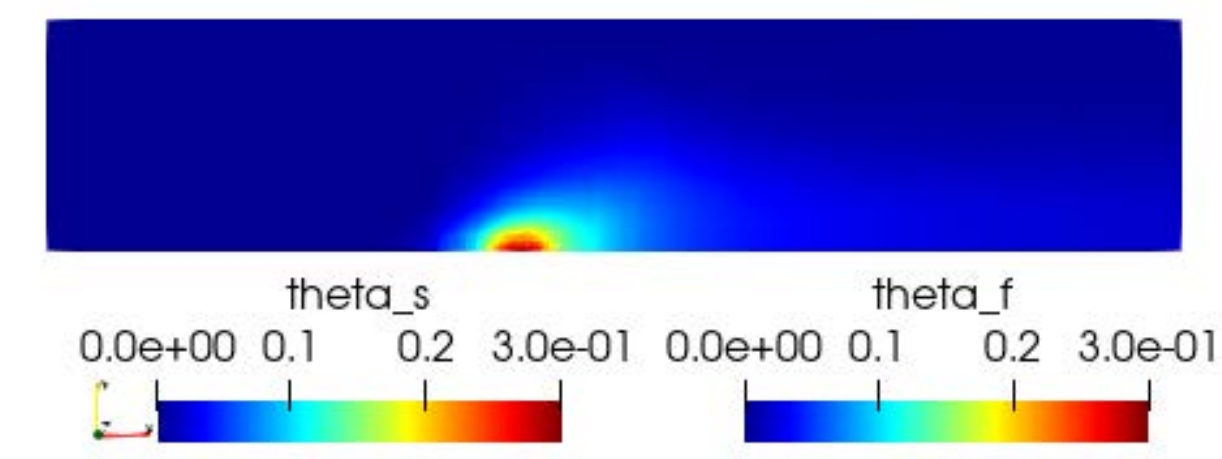}
\caption{Temperature (OpenFOAM)}
\end{subfigure}
\begin{subfigure}{0.32\textwidth}
\includegraphics[width=1\textwidth]{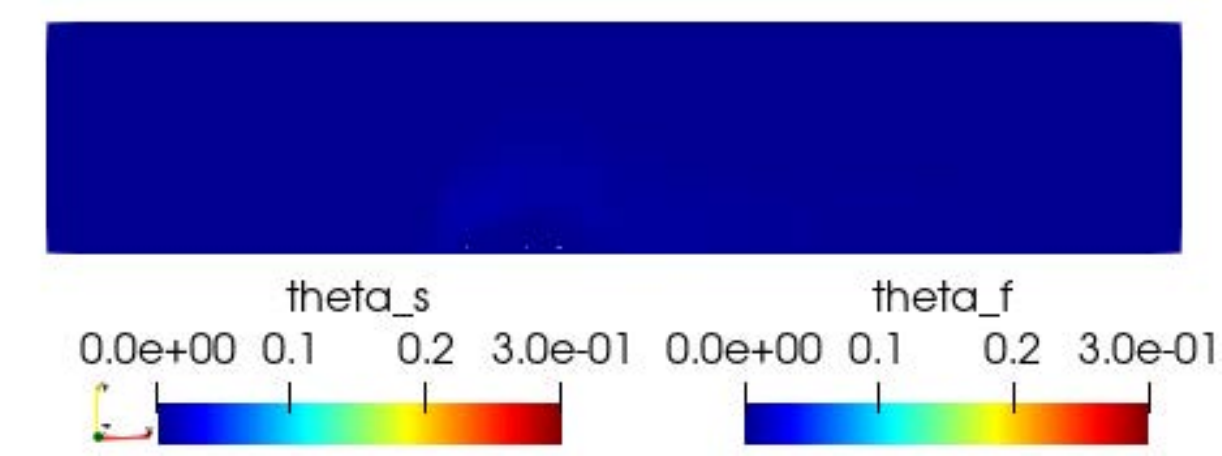}
\caption{Temperature (Difference)}
\end{subfigure}
\caption{\textcolor{black}{A comparison between the SimNet (with modified Fourier feature network) and OpenFoam results for FPGA flow and temperature fields. Results are shown on a 2D slice of the domain.}}
\label{fig:fpga_comparison}
\end{figure}

Using SimNet, we simulate this conjugate heat transfer problem with Fourier feature network, modified Fourier feature network, and SiReN on the full geometry, and also with Fourier feature network on the half geometry using symmetry boundary conditions. Results for the loss curves are reported in Figure \ref{fig:fpga_loss}, and the pressure drop and peak temperature obtained from various runs are reported in Table \ref{table:fpga_results}. Pressure drop and peak temperature are the two quantities that are critical for the heat sink design optimization. The Fourier and modified Fourier feature networks show better convergence behavior compared to the SiReNs as shown in figure \ref{fig:fpga_loss}. \textcolor{black}{This figure also includes the flow convergence results for a Fourier feature model without SDF loss weighting and a standard fully connected model, showing that they fail to provide a reasonable convergence and highlighting the importance of SDF loss weighting and more advanced architectures}. The streamlines and temperature profile obtained from the SimNet model with modified Fourier feature network are also shown in Figure \ref{fig:fpga_streamlines}. \textcolor{black}{A comparison between the SimNet and OpenFoam results for flow and temperature fields is also presented in Figure \ref{fig:fpga_comparison}.}

\begin{table}[h]
\begin{center}
\caption{\label{table:fpga_results} A comparison for the pressure drop and peak temperature obtained from various models.}
\begin{tabular}{ l | c c } 
 \hline
 Case Description & $P_{drop}$ $(Pa)$ & $T_{peak}$ (\textdegree{}$C$) \\
 \hline
 SimNet: Fourier network (axis spectrum) & 25.47 & 73.01\\

 SimNet: Fourier network (partial spectrum) with symmetry &  29.03  &  72.36 \\

  SimNet: Modified Fourier network &  29.17  &  72.52 \\

 SimNet: SiReN &  29.70  &  72.00 \\
\hline
 OpenFOAM Solver & 27.82 & 56.54\\

 Commercial Solver & 24.04 & 72.44\\
 \hline
\end{tabular}
\end{center}
\end{table}

\subsection{Blood flow in an Intracranial Aneurysm} \label{example2}
We demonstrate the ability of SimNet to work with STL geometries from a CAD system. Using the SimNet's TG module, we simulate the flow inside a patient specific geometry of an aneurysm depicted in Figure \ref{subfig:aneurysm_geometry}. Problem setup details are provided in Appendix \ref{appendix:anneurysm}. The SimNet results for the distribution of velocity magnitude and pressure developed inside the aneurysm are shown in Figures \ref{subfig:aneurysm_velocity} and  \ref{subfig:aneurysm_pressure}, respectively. Using the same geometry, the authors in \cite{raissi2020hidden} solve this as an inverse problem using concentration data from the spectral/hp-element solver Nektar. We solve this problem as a forward problem without any data. When solving the forward CFD problem with non-trivial geometries, one of the key challenges is getting the flow to develop correctly, especially inside the aneurysm sac. The streamline plot in Figure \ref{subfig:aneurysm_streamline} shows that SimNet successfully captures the flow field very accurately. 
 
\begin{figure}[htp]
\centering
\begin{subfigure}{0.18\textwidth}
\includegraphics[width=1\textwidth]{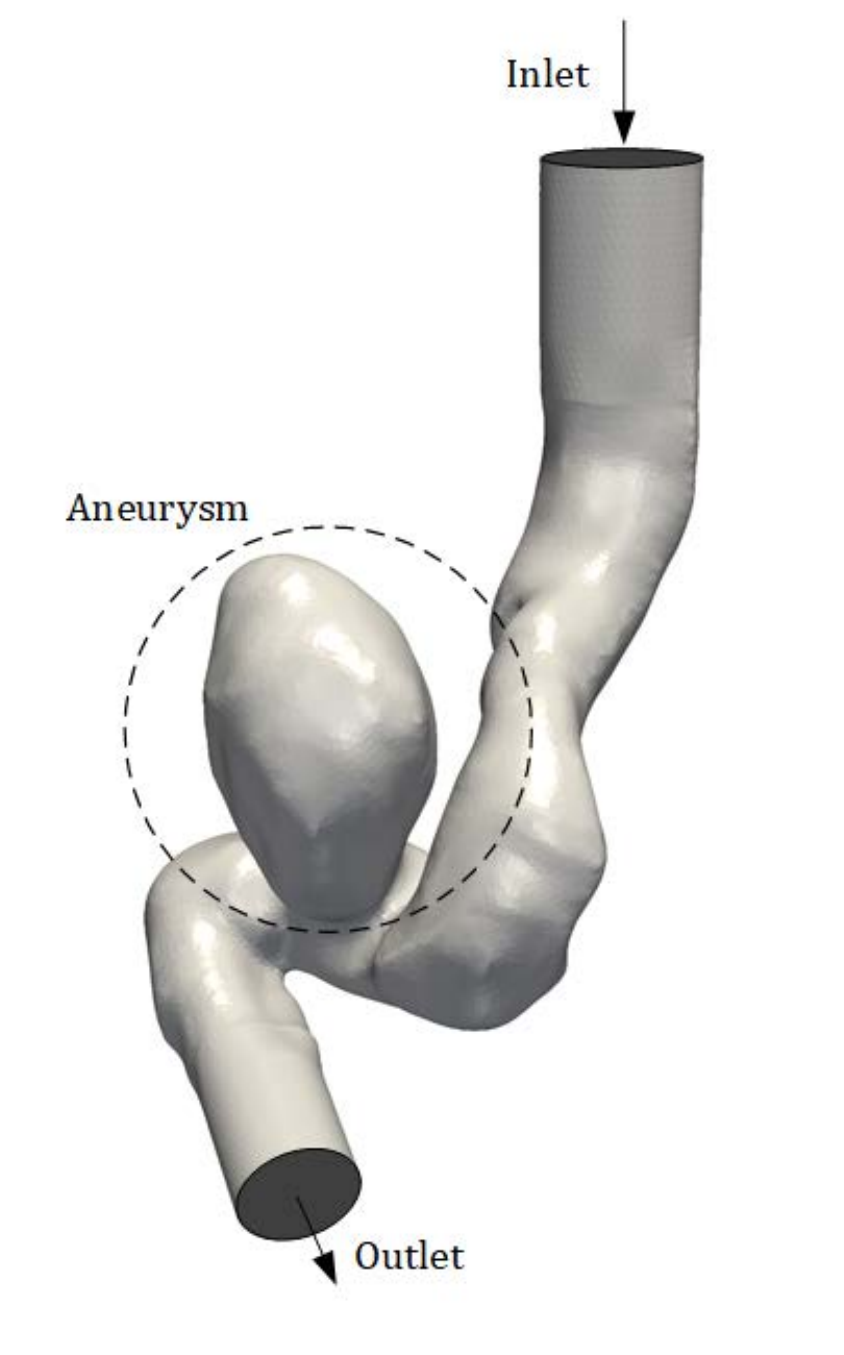}
\caption{Aneurysm geometry}
\label{subfig:aneurysm_geometry}
\end{subfigure} 
\hspace{0.05\textwidth}
\begin{subfigure}{0.31\textwidth}
\includegraphics[width=1\textwidth]{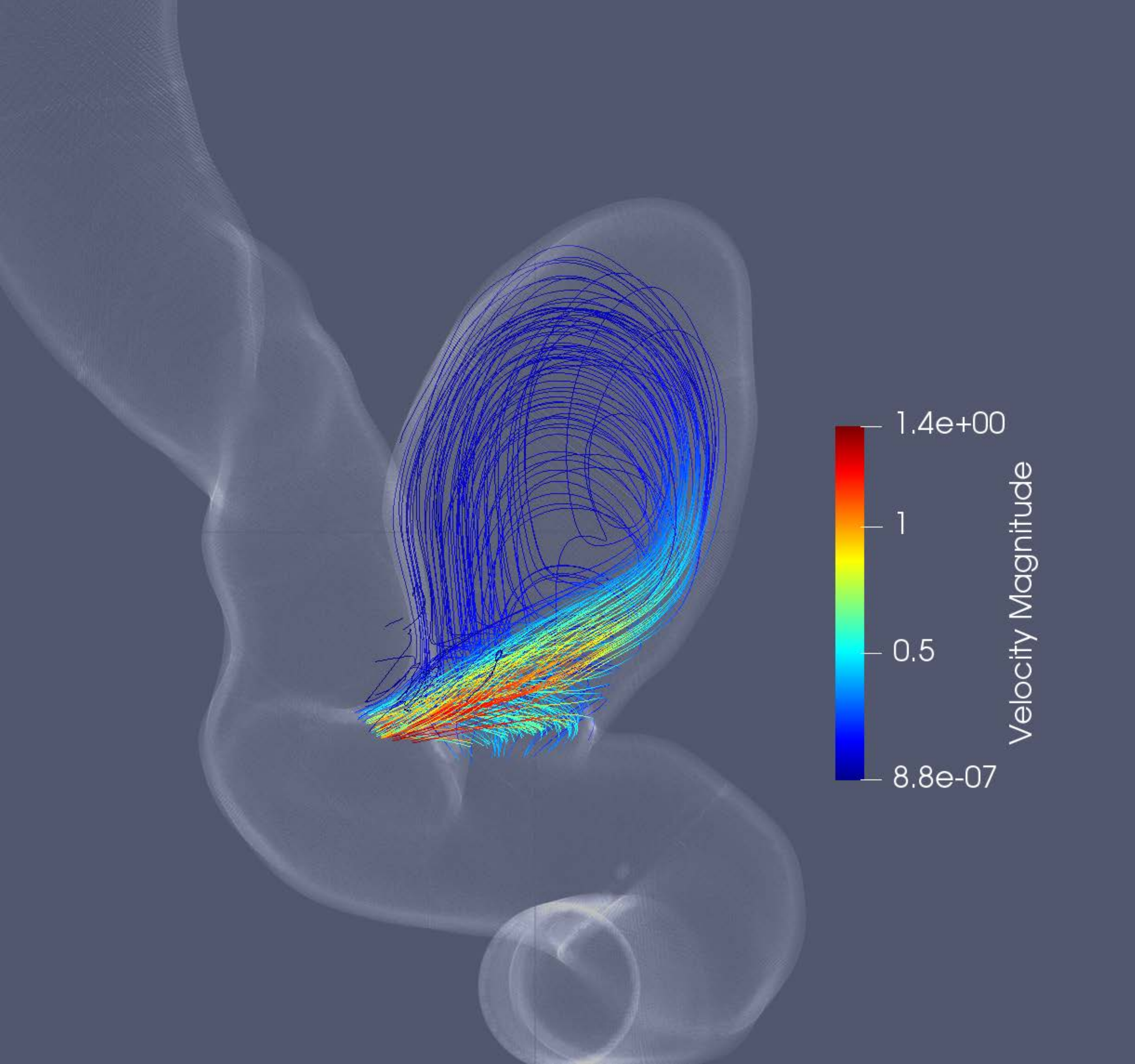}
\caption{Streamlines inside the aneurysm sac}
\label{subfig:aneurysm_streamline}
\end{subfigure}
\\
\begin{subfigure}{0.495\textwidth}
\includegraphics[width=1\textwidth]{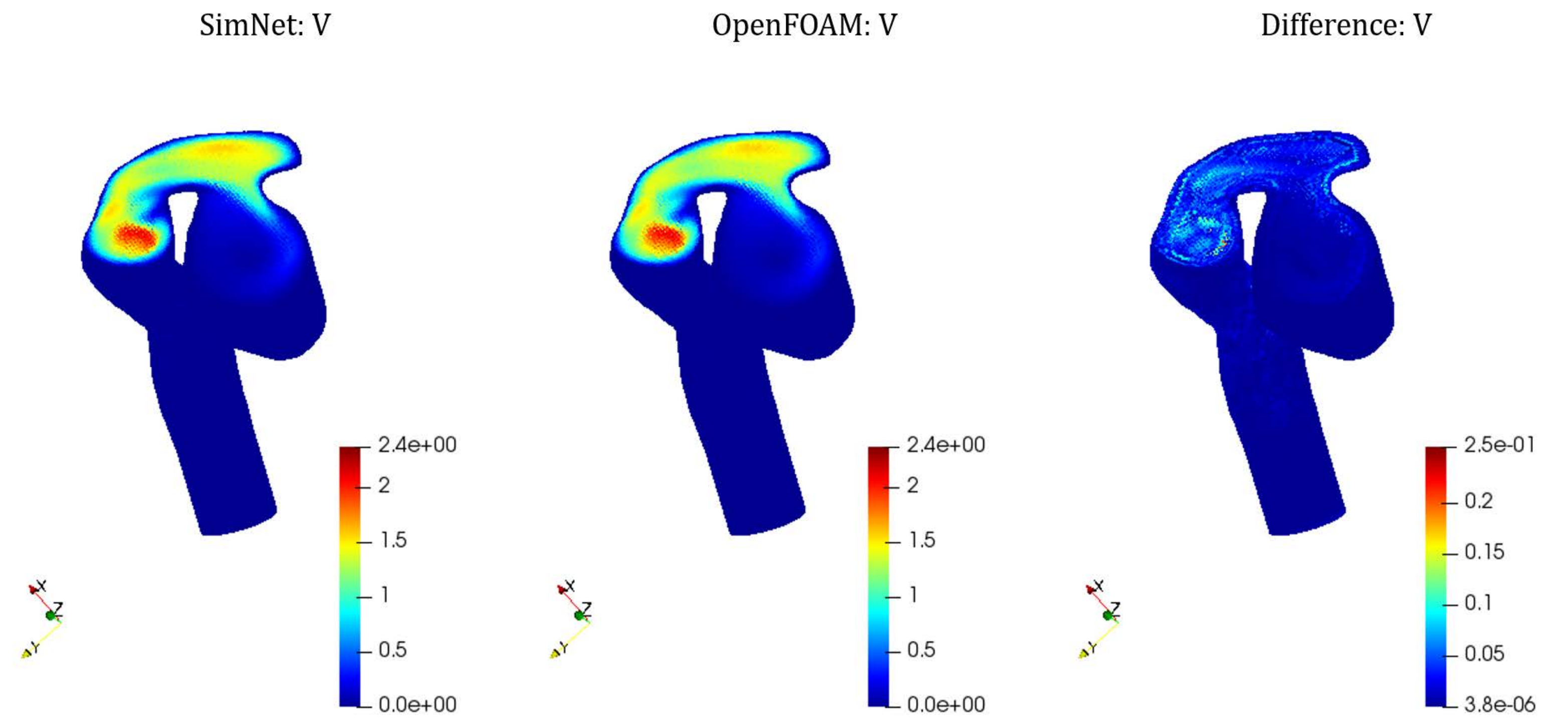}
\caption{Velocity magnitude comparison}
\label{subfig:aneurysm_velocity}
\end{subfigure} 
\begin{subfigure}{0.495\textwidth}
\includegraphics[width=1\textwidth]{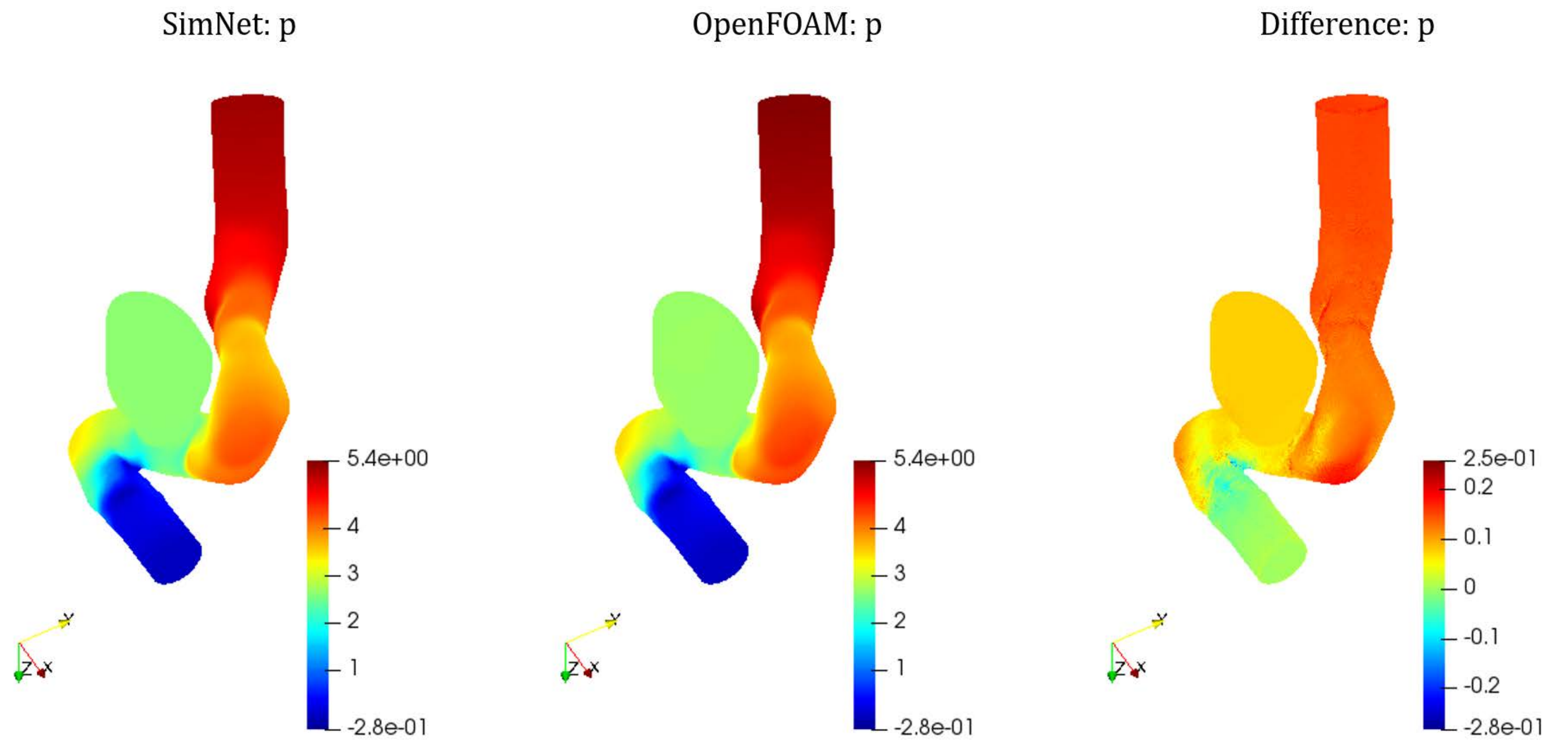}
\caption{Pressure comparison}
\label{subfig:aneurysm_pressure}
\end{subfigure}
\caption{SimNet results for the aneurysm problem, and a comparison between the SimNet and OpenFOAM results for the velocity magnitude and pressure.}
\label{fig:aneurysm}
\end{figure}

\subsection{Design optimization for multi-physics industrial systems} \label{example3}
\textcolor{black}{In this example, we show that SimNet can solve several, simultaneous design configurations in a multi-physics, design space exploration problem much more efficiently than traditional solvers. This is possible because unlike a traditional solver, a neural network trains with multiple design parameters in a single training run. Forward solution of parameterized, complex geometry with turbulent fluid flow between thinly spaced fins and no training data makes this problem extremely challenging for the neural networks. Once the training is complete, several geometry or physical parameter combinations can be evaluated using inference as a post-processing step, without solving the forward problem again. Such  throughput enables more efficient design optimization and design space exploration tasks for complex systems in science and engineering.}

\begin{figure}[htp]
\centering
\begin{subfigure}{0.23\textwidth}
\includegraphics[width=1\textwidth]{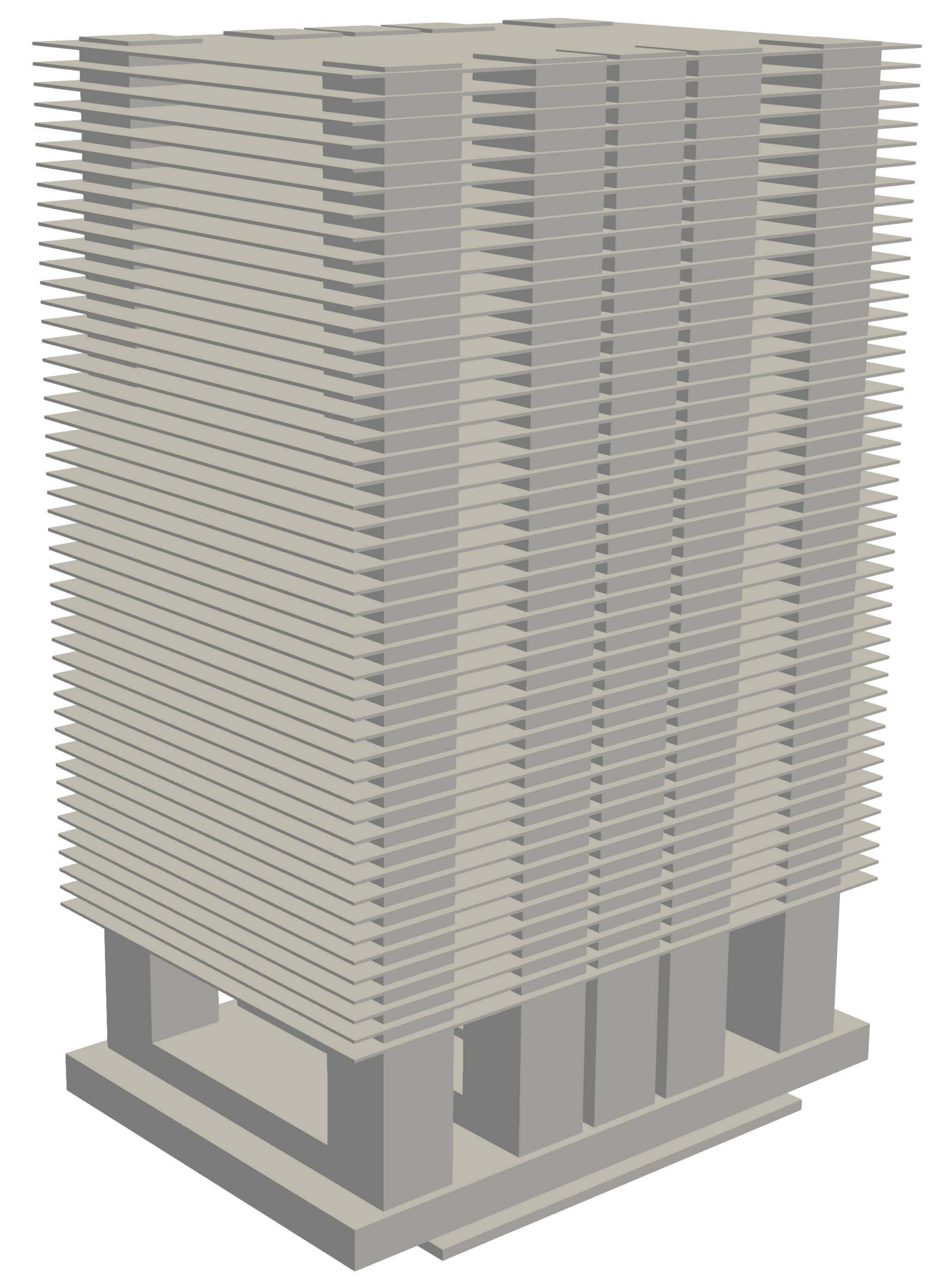}
\caption{NVSwitch base geometry}
\label{fig:3fin_geometry}
\end{subfigure} 
\hspace{0.02\textwidth}
\begin{subfigure}{0.73\textwidth}
\includegraphics[width=1\textwidth]{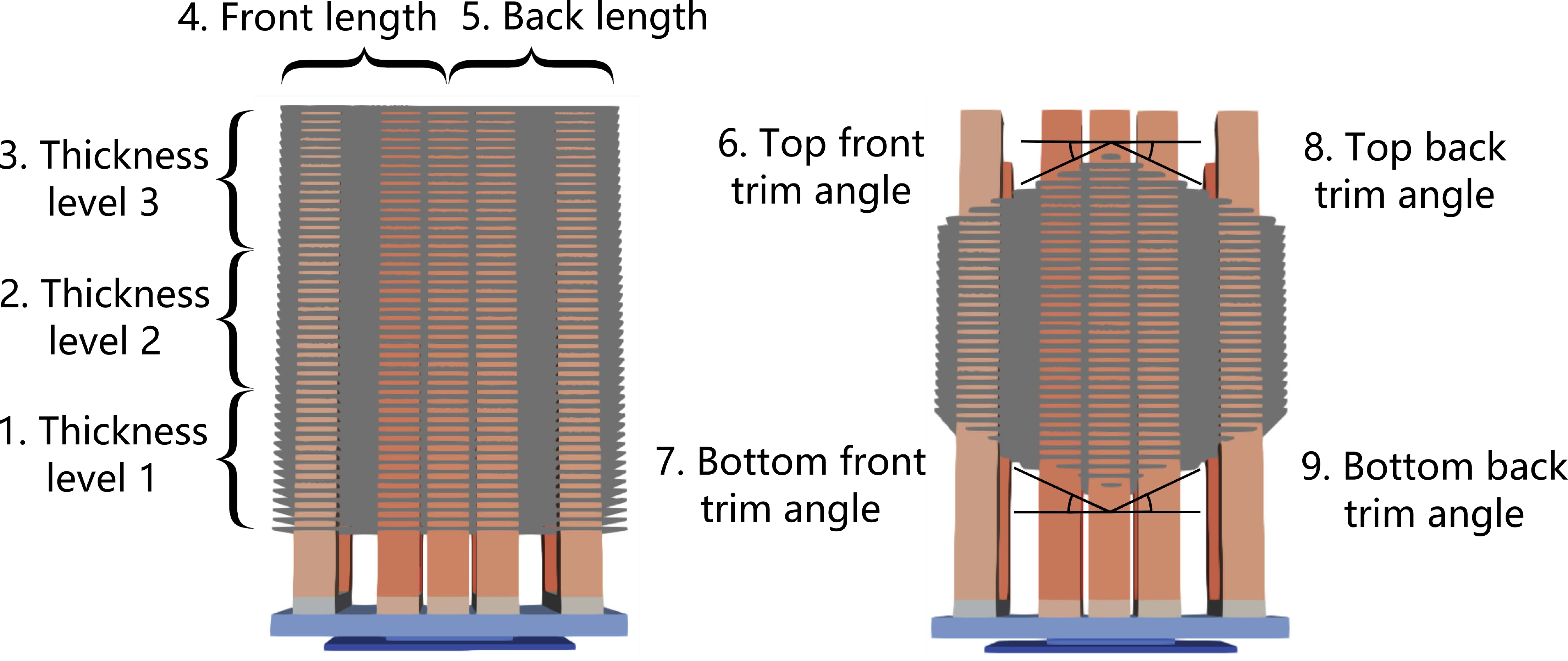}
\caption{NVSwitch design parameters}
\label{fig:3fin_results}
\end{subfigure}
\caption{NVSwitch base geometry and design parameters.}
\label{fig:limerock}
\end{figure}

\textcolor{black}{Here, we train a conjugate heat transfer problem over the Nvidia's NVSwitch heat sink whose fin geometry are variable, as shown in Figure \ref{fig:limerock} (nine geometry variables in total). Following the training, we perform a design optimization to find out the most optimal fin configuration. In heat sink design, usually the objective is to minimize the peak temperature that can be reached at the source chip while satisfying a maximum pressure drop constraint. This is necessary to meet the operating temperature requirements of the chip on which the heat sink is mounted for cooling purposes. For details on the problem setup, refer to Appendix \ref{appendix:3fin}.} 

\begin{figure}[htp]
    \centering
    \includegraphics[width=0.75\textwidth]{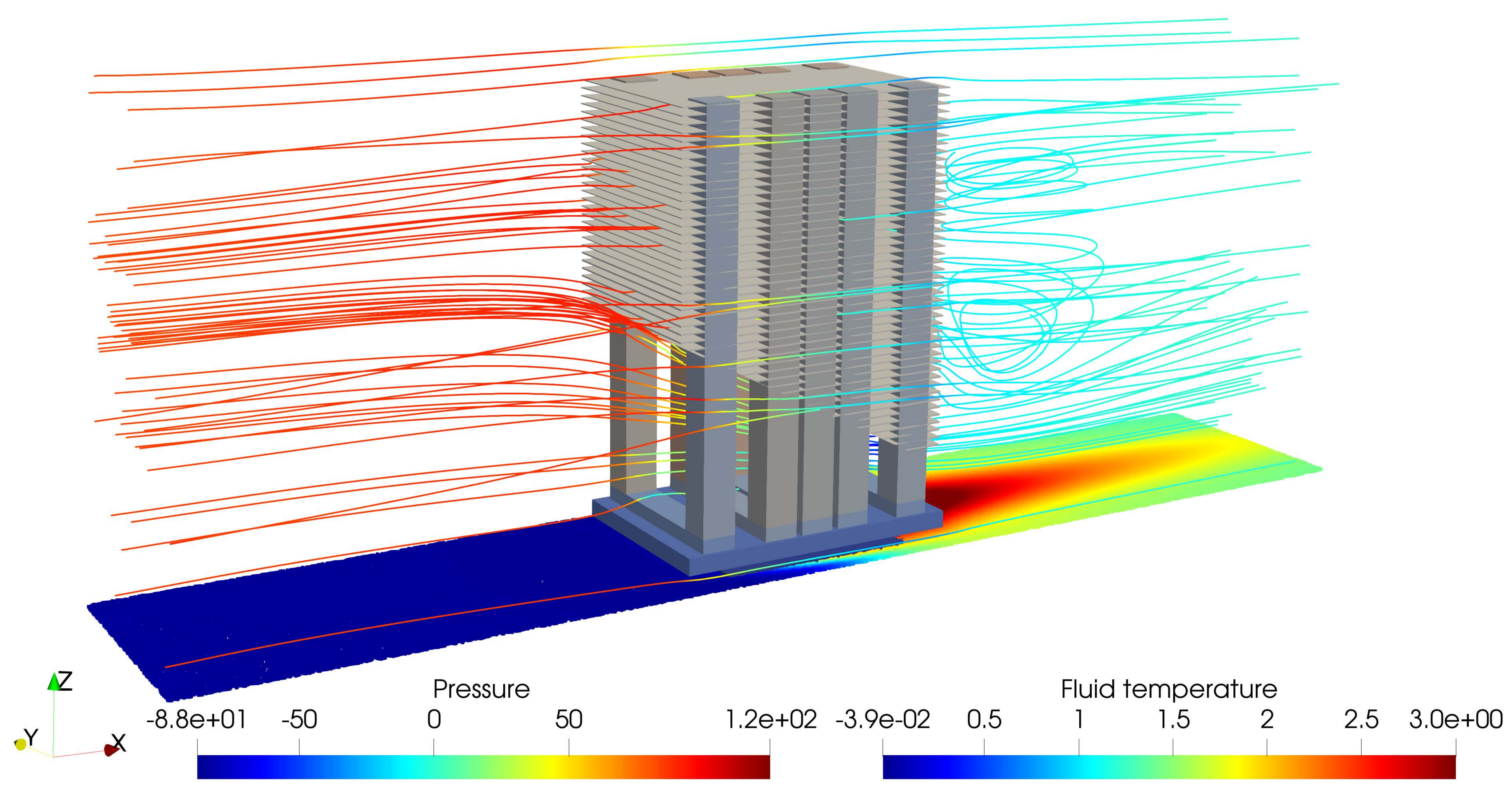}
    \caption{Streamlines colored with pressure and temperature profile in the fluid for optimal NVSwitch geometry.}
    \label{fig:limerock_streamlines}
\end{figure}

\textcolor{black}{The fluid and heat neural networks in this example consist of 12 variables, i.e. three spatial variables and nine geometry parameter variables. Using SimNet, we train these two parameterized neural networks, and then use the trained models to compute the pressure drops and peak temperatures corresponding various combinations using the minimum, maximum and median of these 12 parameters, resulting in 3**12 = 531,441 different heat sink designs. The optimized design is the one that minimizes the peak temperature while satisfying a pressure drop constraint as the optimal design. Figure \ref{fig:limerock_streamlines} shows the streamlines and temperature profile for the optimal NVSwitch geometry. Details of this geometry are reported in Appendix \ref{appendix:3fin}.}

\textcolor{black}{To confirm the accuracy of the parameterized models, we take the NVSwitch base geometry and compare the SimNet results (obtained from the parameterized model) for pressure drop and peak temperature with the OpenFOAM and commercial solver results, reported in Table \ref{table:parameterized1}.}. OpenFOAM over predicts the commercial solver results by 4.5\% while the SimNet results under predict the commercial solver result by about 15\%. 

\begin{table}[h]
\begin{center}
\caption{\label{table:parameterized1}A comparison for the solver and SimNet results for NVSwitch pressure drop and peak temperature.}
\begin{adjustbox}{width=\columnwidth,center}
\begin{tabular}{ l|c c c  } 
 \hline
 Property                   & OpenFOAM Single Run & Commercial Solver Single Run & SimNet Parameterized Run \\
 \hline
 Pressure Drop $(Pa)$           & $133.96$  & $128.30$ &  $109.53$ \\
 Peak Temperature $(^{\circ} C)$ & $41.55$ &  $43.57$ &   $39.33$\\
\hline
\end{tabular}
\end{adjustbox}
\end{center}
\end{table}

\textcolor{black}{By parameterizing the geometry, SimNet significantly accelerates design optimization when compared to traditional solvers, which are limited to single geometry simulations. The total compute time required by OpenFOAM, a commercial solver, and SimNet for this design optimization task is reported in Table \ref{table:parameterized_time}. The OpenFOAM and commercial solver runs are run on 22 CPU processors, and the SimNet runs are on 8 V100 GPUs. It can be seen that SimNet can solve this design optimization problem significantly faster than traditional solvers. Increasing the number of design variables or the number of designs to be evaluated magnifies the difference in the time taken for the two approaches of neural network and traditional solvers since for the neural network, once the model is already trained, each of these computations during inference phase take only a fraction of a second. SimNet accelerates the simulation by 45,000x compared to the commercial solver and by 135,000x compared to OpenFOAM.} 

\begin{table}[h]
\begin{center}
\caption{\label{table:parameterized_time} Total compute time needed for different solvers for the NVSwitch heat sink design optimization}
\begin{tabular}{ l|c c c c  } 
 \hline
 Solver                     & OpenFOAM & Commercial Solver & SimNet \\
 \hline
 Compute Time (x 1000 hrs.)       & $405935$  & $137494$ & $3$ \\
\hline
\end{tabular}
\end{center}
\end{table}

\subsection{Inverse problems} \label{example4}
Many applications in science and engineering involve inferring unknown system characteristics given measured data from sensors or imaging for certain dependent variables describing the behavior of the system. Such problems usually involve solving for the latent physics using the PDEs as well as the data. This is done in SimNet by combining the data with PDEs to decipher the underlying physics.  

\begin{figure}[htp]
    \centering
    \includegraphics[width=0.5\textwidth]{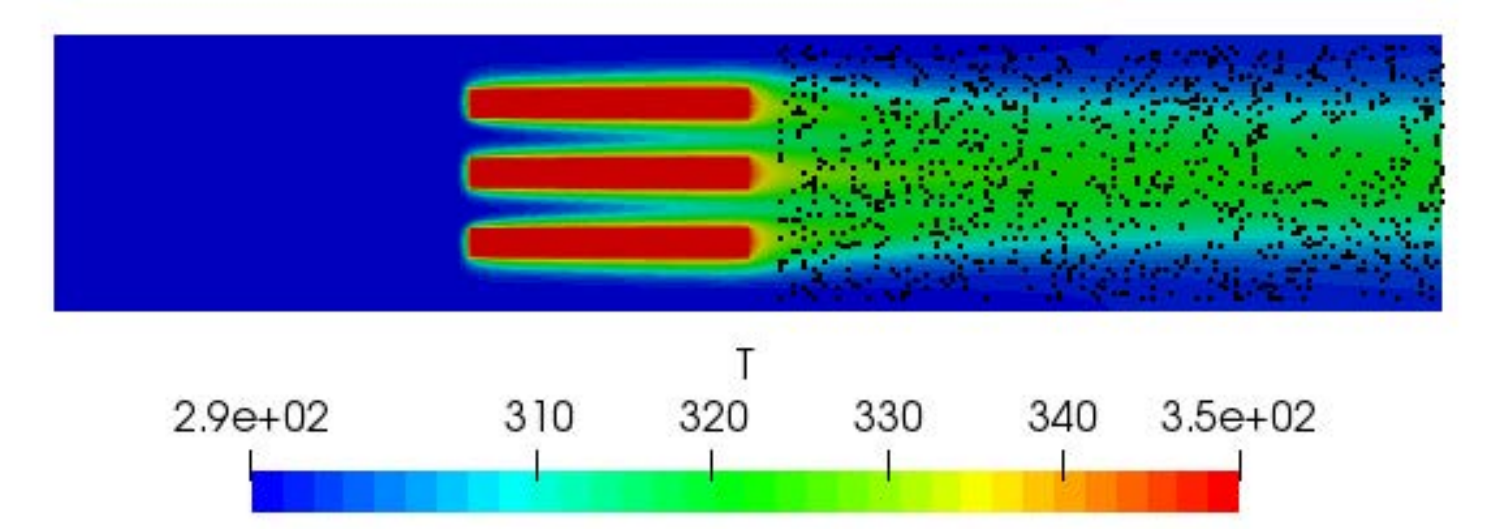}
    \caption{A batch of training points sampled from the OpenFOAM data.}
    \label{fig:inverse}
\end{figure}

Here, we demonstrate the ability of SimNet to solve data assimilation and inverse problems on a 2D cross-section of the 3-fin heat sink example. Given the data consisting of flow velocities, pressure and temperature, all of which that can be measured, the task is to infer the flow characteristics such as flow viscosity and thermal diffusivity. In reality, the data is collected using measurements but for the purpose of this example, synthetic data generated by OpenFOAM is used. As the majority of diffusion of temperature occurs in the wake of the heat sink (Figure \ref{fig:inverse}), we sample the training points only from this wake zone.

\begin{figure}[htp]
\centering
\begin{subfigure}{0.33\textwidth}
\includegraphics[width=1\textwidth]{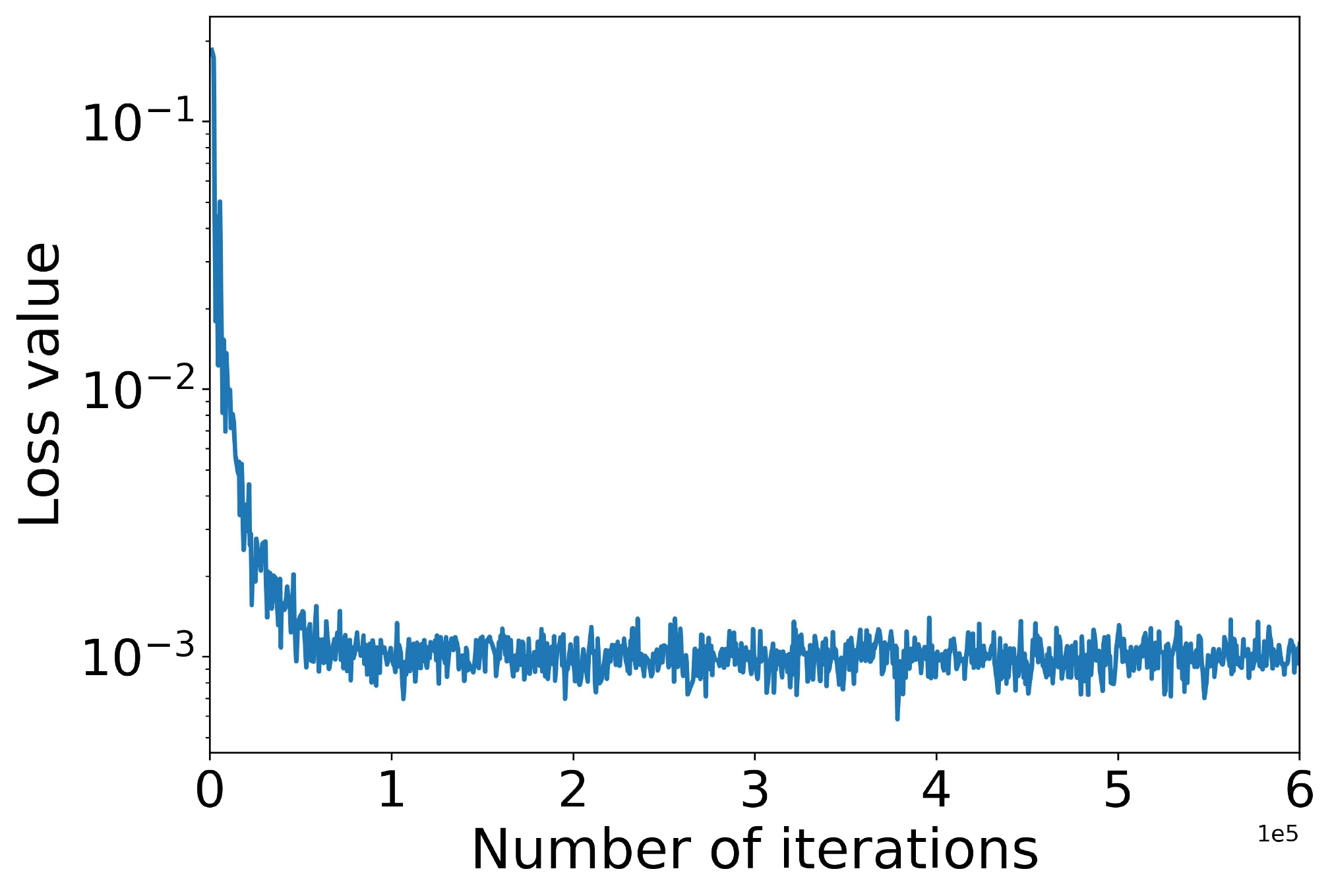}
\caption{Inverse 2D 3-fin training}
\end{subfigure} 
\begin{subfigure}{0.33\textwidth}
\includegraphics[width=1\textwidth]{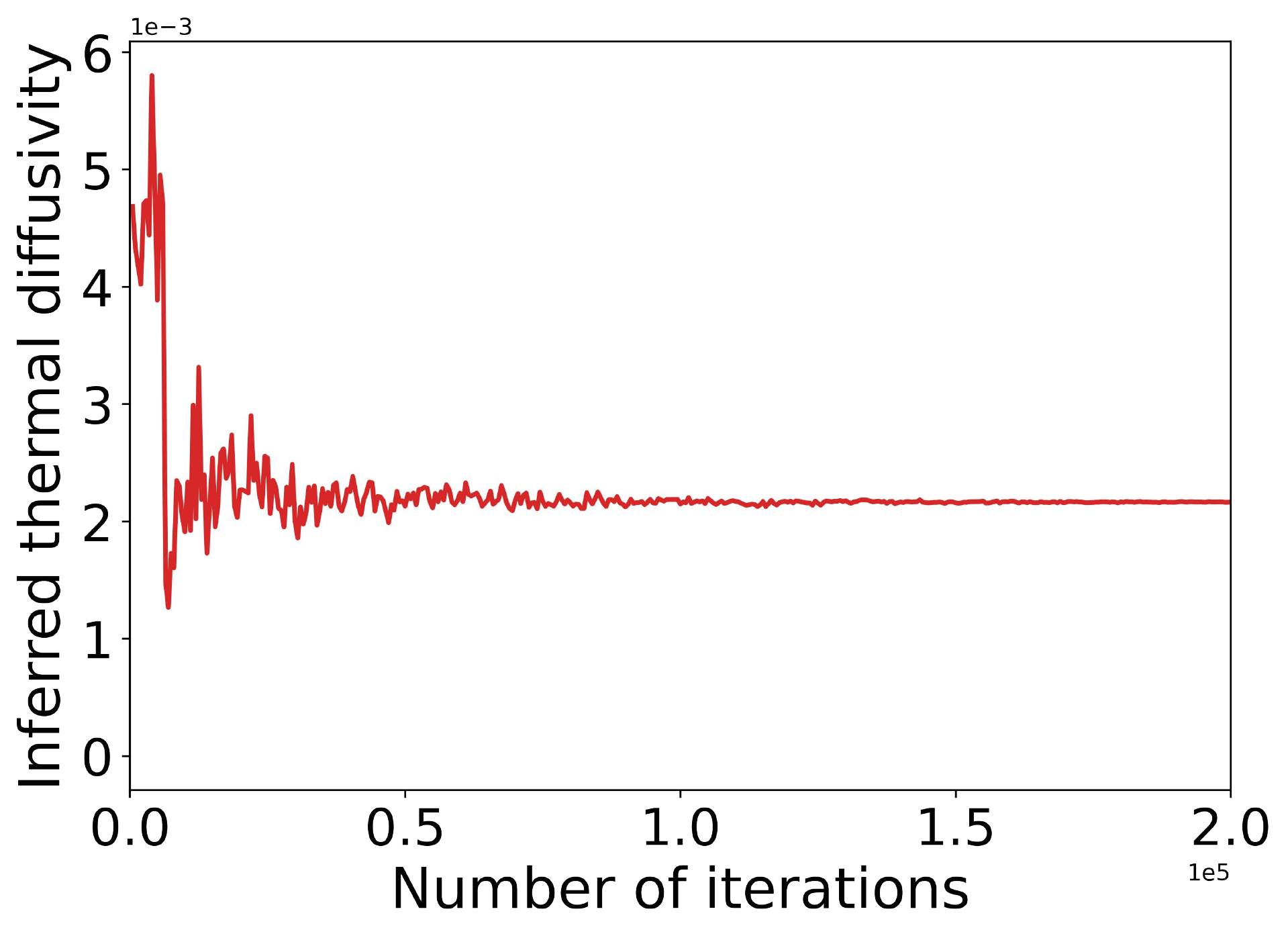}
\caption{Inferred thermal diffusivity}
\end{subfigure}
\begin{subfigure}{0.33\textwidth}
\includegraphics[width=1\textwidth]{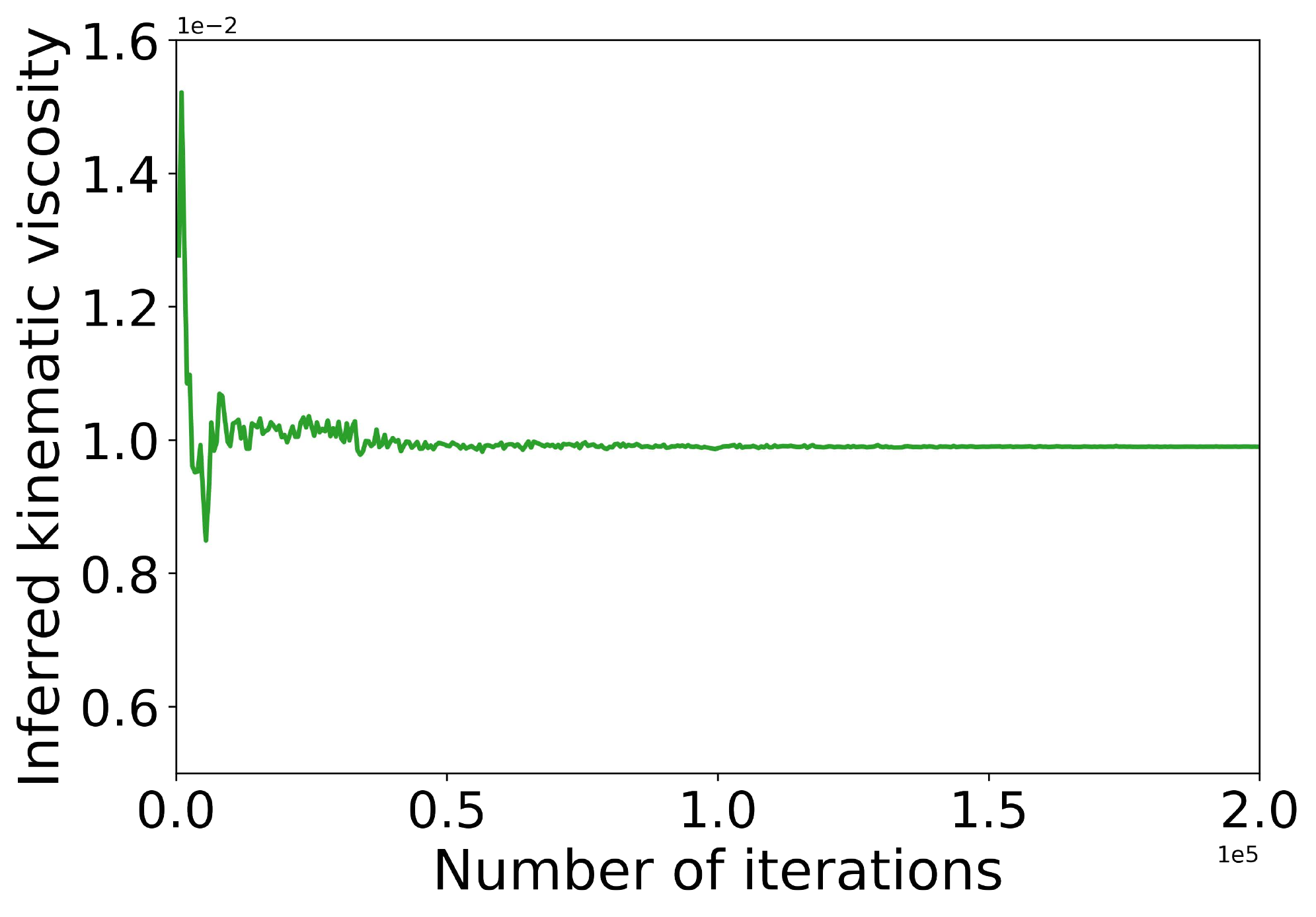}
\caption{Inferred kinematic viscosity}
\end{subfigure}
\caption{Solution to the 2D 3-fin inverse problem using SimNet.}
\label{fig:inverse_loss}
\end{figure}

Next, we construct a neural network model with a hybrid data and physics-driven loss function. Specifically, we require the neural network outputs (i.e. $u$, $v$, $p$, and $T$) to fit to the measurements, and also satisfy the governing laws of the system that includes the Navier-Stokes and advection-diffusion equations. Here, the quantities of interest (i.e., flow viscosity and thermal diffusivity) are also modeled as trainable variables, and are inferred by minimizing the hybrid loss function as shown in Figure \ref{fig:inverse_loss}. A comparison between the predicted SimNet values and the ground truth for flow viscosity and thermal diffusivity is reported in Table \ref{table:inverse}.

\begin{table}[h]
\begin{center}
\caption{\label{table:inverse} A comparison between the ground truth and the inferred flow characteristics.}
\begin{tabular}{ l|c c  } 
 \hline
 Property & OpenFOAM (Ground truth) & SimNet (Predicted) \\
 \hline
 Kinematic viscosity $(m^2/s)$      & $1.00e-2$      & $9.90e-3$ \\ 
 Thermal diffusivity $(m^2/s)$      & $2.00e-3$      & $2.16e-3$ \\
 \hline
\end{tabular}
\\
\end{center}
\end{table}

\section{Performance Upgrades and Multi-GPU Training} \label{HPC}

\subsection{XLA compiler}
XLA (Accelerated Linear Algebra) is a domain specific compiler that allows for just-in-time
compilation of TensorFlow graphs. Neural network solvers implemented in SimNet may have many peculiarities including the presence of many point-wise operations. Such operations, while being computationally inexpensive, heavily use the memory subsystem of a GPU. XLA allows for kernel fusion, so that many of these operations can be computed simultaneously in a single kernel and thereby reducing the number of memory transfers from GPU memory to the compute units. Based on our experiments on the FPGA problem, kernel fusion using XLA accelerates a single training iteration in SimNet by up to 3.3x.

\subsection{Learning rate scaling in multi-GPU training}
An effective way of reducing the time to convergence is to parallelize the training process across multiple GPUs. The most common multi-GPU parallelization strategy is data parallelism where a given global training batch is split into multiple sub-batches for each GPU. Each GPU then performs the forward and backward passes for its sub-batch and the gradients are accumulated across all the GPUs using an allreduce algorithm. This form of data parallelism is the most computationally efficient when the batch size per GPU is kept constant instead of the global batch size.

\begin{figure}[h]
\centering
\begin{subfigure}{0.495\textwidth}
\includegraphics[width=1\textwidth]{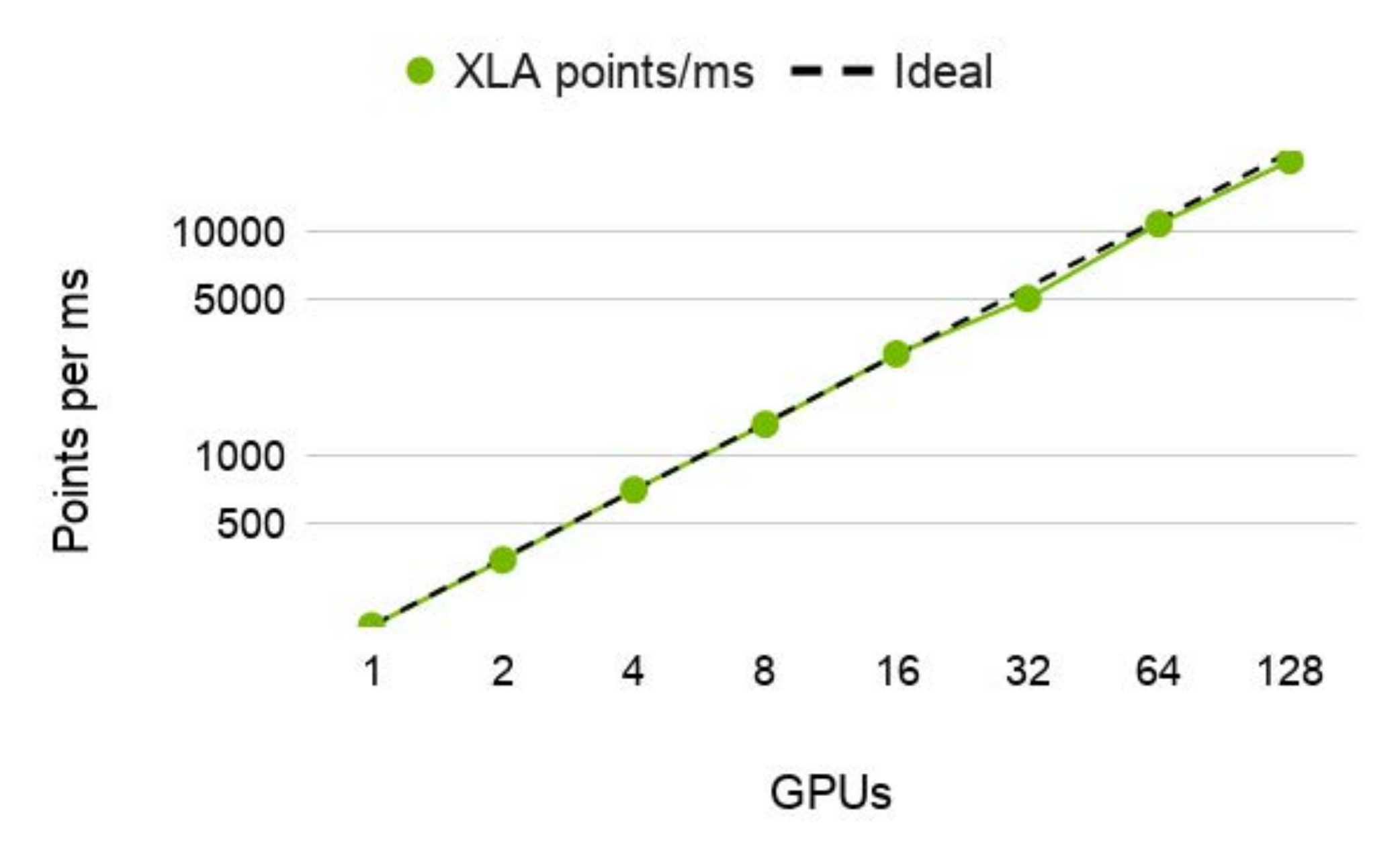}
\caption{Points per ms weak scaling}
\end{subfigure} 
\begin{subfigure}{0.495\textwidth}
\includegraphics[width=1\textwidth]{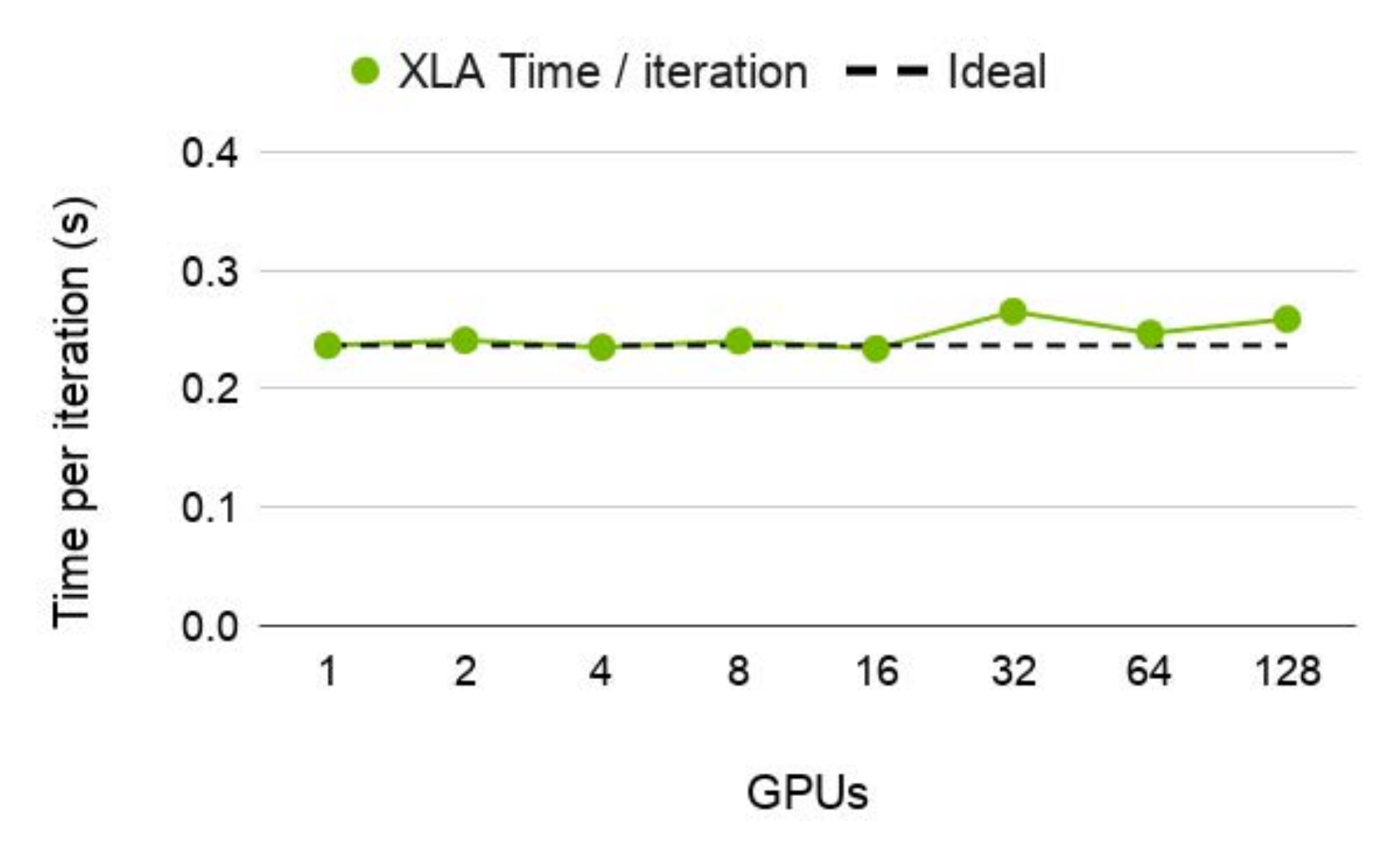}
\caption{Time per iteration weak scaling}
\end{subfigure}
\caption{Weak Scaling on V100 GPUs.}
\label{fig:multiGPU1}
\end{figure}

It was shown in \cite{goyal2017accurate} that the total time to convergence can be reduced linearly with the number of GPUs by proportionally increasing the learning rate. However, doing that naively would cause the model to diverge since the initial learning rate can be very high. An effective solution for this is to have an initial warmup period when the learning rate gradually increases from the baseline to the scaled learning rate. SimNet implements a constant and a linear learning rate warmup scheme in conjunction with an exponential decay baseline learning rate schedule, and the details are reported in Appendix \ref{appendix:warmup}. By running a multi-GPU training, larger batch sizes can be used allowing larger models to be run without increasing the time as shown in Figure \ref{fig:multiGPU1}a. Doing so, the time per iteration remains nearly constant, as shown in Figure \ref{fig:multiGPU1}b. For the multi-GPU cases, the learning rate is gradually increased from the baseline case and this allows the model to train without diverging early on and allows the model to converge faster as a result of the increased global batch size coupled with the increased learning rate. Figure \ref{fig:fpga_tf32}a shows the loss function evolution as the number of GPUs is increased from 1 to 16 for the NVSwitch heatsink case. 

\subsection{TF32 math mode}
\begin{figure}[htp]
\centering
\begin{subfigure}{0.475\textwidth}
\includegraphics[width=1\textwidth]{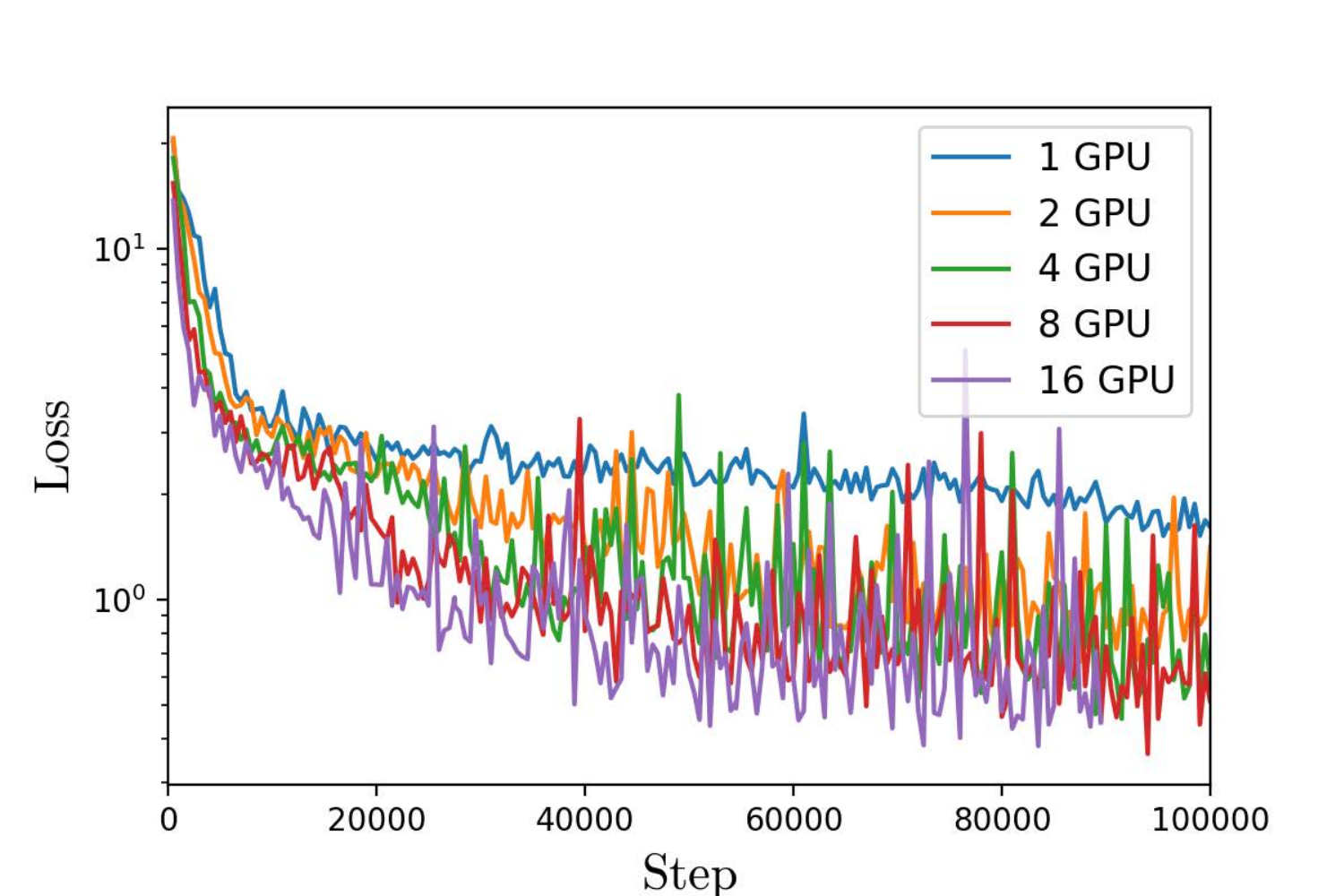}
\caption{Strong Scaling on A100}
\end{subfigure} 
\begin{subfigure}{0.51\textwidth}
\includegraphics[width=1\textwidth]{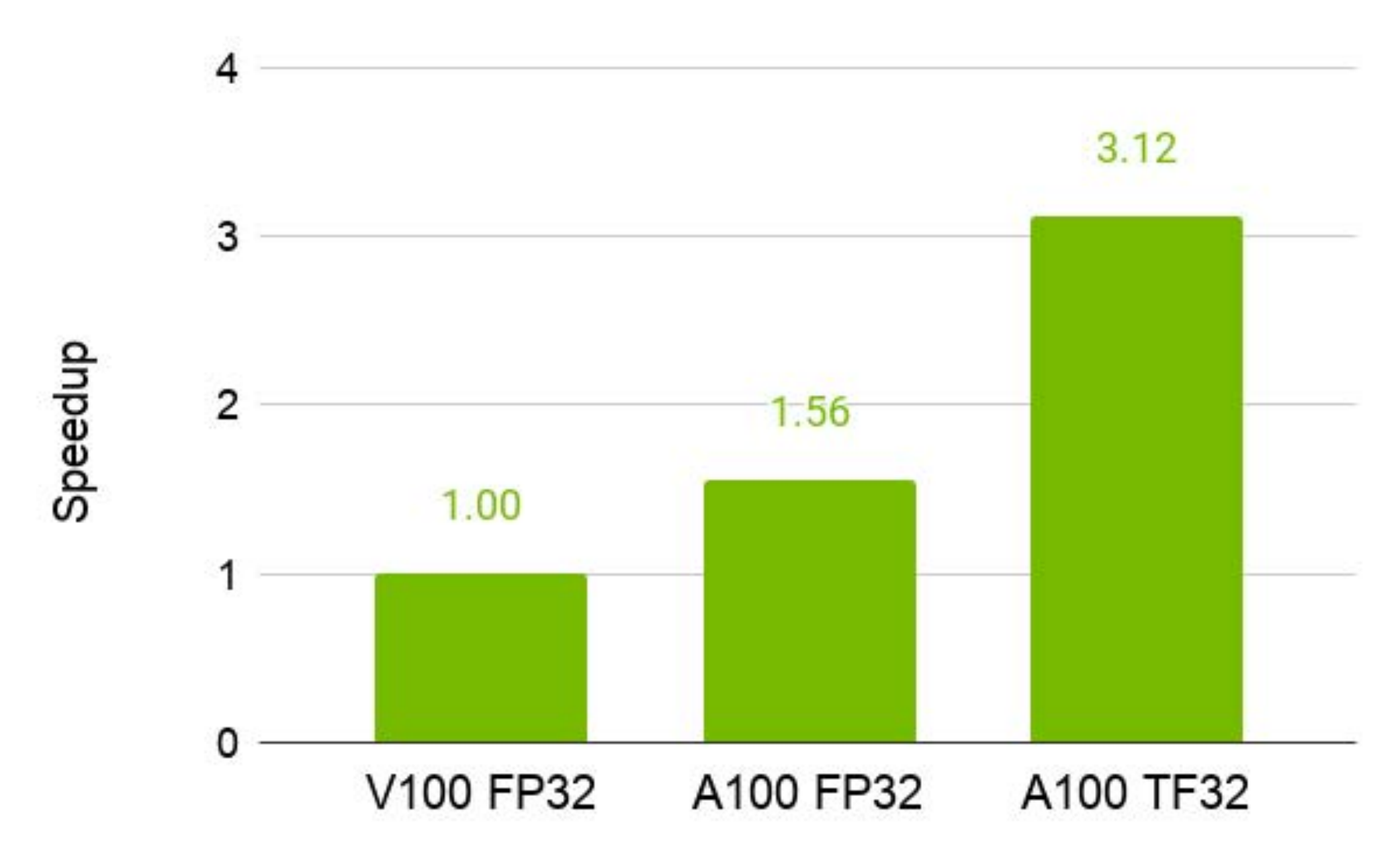}
\caption{FPGA Speedup with A100 GPUs and TF32 Precision}
\end{subfigure}
\caption{Accelerated training on A100 GPUs.}
\label{fig:fpga_tf32}
\end{figure}

TensorFloat-32 (TF32) is a new math mode available on NVIDIA A100 GPUs for handing matrix math and tensor operations used during the training of a neural network. With this feature, and based on our experiments on the FPGA heat sink problem, we can obtain up to 1.6x speed-up over FP32 on A100 GPUs and up to 3.1x speed-up over FP32 on V100 GPUs. This allows us to achieve similar accuracy compared to FP32 at a reduced training time, as shown in Figure \ref{fig:fpga_tf32}b.

\section{Conclusion} \label{sec:conclusion}

SimNet is an end-to-end AI-driven simulation framework with unique, state-of-art architectures that enables rapid training convergence of forward, inverse problems and data assimilation problems for real world geometries and multiple physics types without any training data. SDF is used for loss weighting, which has been shown to significantly improve the training convergence in cases where the geometry has sharp corners and results in sharp gradients in the solution of the differential equation. SimNet's TG module enables the users to import tessellated geometries from CAD programs. For systems governed by the Navier-Stokes equations, conservation of mass or continuity is imposed globally as well as locally  in SimNet to further improve the convergence and accuracy. \textcolor{black}{SimNet enables the neural network solvers to simulate, high Reynolds number turbulent flows for industrial applications. To the authors knowledge, this is the first such application of Physics driven neural networks for turbulent flows.}

SimNet is designed to be flexible so that users can leverage the functionality in the existing toolkit and focus on solving their problem well rather than re-creating the tools. To this end, there are various APIs that enable the user to implement their own equations to simulate the physics, implement their own geometry primitives or importing complex tessellated geometries, or implement a variety of domains/boundary conditions. The geometry parameterization in the CSG module allows the neural network to address the entire range of all given parameters in a single training, as opposed to the traditional simulations that run one at a time. The inference for any design configuration can then be completed in real time. This accelerates the simulation with neural network solvers by orders of magnitude. 

\textcolor{black}{In a broader context, SimNet provides a framework that is capable of addressing major areas across the computational science and engineering. So far, extensive comparisons of SimNet results with open source and commercial solvers show good correlation. However, further applications across a wide range of use cases are required to verify the robustness, accuracy, and applicability of neural network solvers.}

SimNet can be downloaded from: \href{http://developer.nvidia.com/simnet}{developer.nvidia.com/simnet}

\section*{Acknowledgments}
We would like to thank Doris Pan, Anshuman Bhat, Rekha Mukund, Pat Brooks, Gunter Roth, \textcolor{black}{Ingo Wald}, Maziar Raissi and Sukirt Thakur for their assistance and feedback in SimNet development. We also acknowledge Peter Messemer, Mathias Hummel, Tim Biedert and Kees Van Kooten for integration with Omniverse.

\bibliographystyle{unsrt}  
\small
\bibliography{references}

\appendix

\section {FPGA Problem Description}\label{appendix:fpga}

The dimensions of the FPGA geometry are summarized in the Table \ref{table:fpga_geometry}. Compared to the real FPGA dimensions, these dimensions are normalized such that the channel height is set to $1$. The channel walls are treated as adiabatic and the interface boundary conditions are applied at the fluid-solid  interface.

\begin{table}[h]
\begin{center}
\caption{\label{table:fpga_geometry}FPGA Dimensions}
\begin{tabular}{ l|l } 
 \hline
 Dimension  & Value $(m)$  \\
 \hline
 Heat Sink Base ($\ell \times w \times h$)  & $0.65 \times 0.875 \times 0.05$   \\ 
 Fin Dimension ($\ell \times w \times h$) & $0.65 \times 0.0075 \times 0.8625$  \\
 Heat Source ($\ell \times w$) & $0.25 \times 0.25$ \\
 Channel ($\ell \times w \times h$) & $5.0 \times 1.125 \times 1.0$ \\
 \hline
\end{tabular}
\end{center}
\end{table}

\section{Aneurysm Problem Description}\label{appendix:anneurysm}
For the aneurysm simulation, a no-slip boundary condition is applied on the walls of the aneurysm, i.e., $u,v,w=0$. A parabolic flow is used at the inlet, where the flow moves in the normal direction of the inlet and has a peak velocity of $1.5$. The outlet also has a zero pressure condition. The kinematic viscosity of the fluid is set to $0.025$, and the fluid density is a constant and is set to $1.0$. A total number of about 20M training points are used to solve this problem. 

\section{NVSwitch Problem Description}\label{appendix:3fin}
\textcolor{black}{The inlet velocity to the channel is at 5.7 $m/s$. The pressure at the outlet is specified as 0 $Pa$. All the other surfaces of the geometry are treated as no-slip walls. The inlet is at 273.15 $K$. The channel walls are adiabatic. The heat sink has a heat source of 0.0190 x 0.0077 $m^2$ at the bottom of the heat sink situated centrally on the bottom surface. The heat source generates heat such that the temperature gradient on the source surface is 6828.75 $K/m$ in the normal direction. Conjugate heat transfer takes place between the fluid-solid contact surface.}

\textcolor{black}{For this problem, we will vary the fin thickness, fin length at front and back, and fin trim angles, as shown in Figure \ref{fig:limerock}. The ranges of variation for these geometry parameters as well as their optimal values are given in Table \ref{table:limerock_variables}. The maximum allowable pressure drop for this design optimization is assumed to be $103.77 Pa$.}

\begin{table}[htp]
\centering
\caption{\label{table:limerock_variables}Range of variability for the NVSwitch fin design variables and their optimal values}
\begin{adjustbox}{width=\columnwidth,center}
\begin{tabular}{ c|c c c c c c c c c } 
 \hline
 Design & Thickness & Thickness & Thickness & Front & Back & Top front & Bottom front & Top back & Bottom back  \\
 variable & level 1 & level 2 & level 3 & length & length & trim angle & trim angle & trim angle & trim angle \\
\hline
 Range & $(0.0025, 0.0075)$ & $(0.0025, 0.0075)$ & $(0.0025, 0.0075)$ & $(0.5325, 0.6075)$ & $(0.5325, 0.6075)$ & $(0, \pi /6)$ & $(0,\pi/6)$ & $(0,\pi/6)$ & $(0,\pi/6)$\\
 \hline
 Optimal value & $0.0067$ & $0.0061$ & $0.0038$ & $0.5951$ & $0.5414$ & $0.0$ & $0.15 \pi$ & $0.0$ & $0.0$ \\
 \hline
\end{tabular}
\end{adjustbox}
\end{table}

\section{Learning Rate Linear Warm-up} \label{appendix:warmup}
For the linear warmup scheme, the learning rate $\eta$ at step $s$ when run with $n_g$ GPUs is given by
\begin{equation}
    \eta = \min \left\{ \eta_w(n_g), \eta_b \right\},
\end{equation}
where
$\eta_{b}$ is the baseline exponential decay learning rate schedule given by
\begin{equation}
    \eta_b = \eta_{0,b} r^{s/s_d} + \eta_{1,b},
\end{equation}
and $\eta_w(n_g)$ is the warmup learning rate given by
\begin{equation}
    \eta_w(n_g) = \eta_{0,b} \left[ 1 + \left( n-1 \right) \frac{s}{s_w} \right].
\end{equation}
Here, $s_d$ is the baseline learning rate decay steps, $r$ is the decay rate, $\eta_{0,b}$ and $\eta_{1,b}$ determine the start and end learning rates, and $s_w$ is the number of warmup steps.

\end{document}